\documentclass[conference]{IEEEtran}
\usepackage{cite}
\usepackage{url}
\usepackage{amsmath}

\usepackage[ruled,vlined]{algorithm2e}
\SetKwComment{Comment}{$\triangleright$\ }{}
\usepackage{tabularx}
\usepackage{booktabs}
\usepackage{algpseudocode}
\usepackage{lipsum}
\usepackage{xcolor}
\usepackage{xspace}
\usepackage{tcolorbox}
\usepackage{multirow}
\usepackage{graphicx}
\usepackage{subcaption}

\usepackage{etoolbox}

\renewcommand{\paragraph}[1]{\noindent\textbf{\textit{#1.}}}

\definecolor{MyTitleBackColor}{RGB}{255,220,180}
\definecolor{MyBoxBackColor}{RGB}{255,250,240} 

\newtcolorbox{promptbox}[1]{
    colback=MyBoxBackColor,       
    colframe=MyTitleBackColor,    
    colbacktitle=MyTitleBackColor, 
    coltitle=black,                
    fonttitle=\bfseries,           
    arc=2mm,                       
    boxrule=0.8pt,                 
    boxsep=2mm,                    
    top=2mm, bottom=2mm, left=2mm, right=2mm, 
    title=#1,                      
}

\makeatletter
\def\IEEEauthorrefmark#1{\textsuperscript{#1}}
\makeatother

\newcommand{\myname}{\textsc{AURA}\xspace}
\begin{document}
\title{Making Theft Useless: Adulteration-Based Protection of Proprietary Knowledge Graphs in GraphRAG Systems}
\author{
\IEEEauthorblockN{Weijie Wang}
\IEEEauthorblockA{Institute of Information Engineering,\\ Chinese Academy of Sciences\\
China\\
wangweijie@iie.ac.cn\\
~}
\and
\IEEEauthorblockN{Peizhuo Lv}
\IEEEauthorblockA{Department of Computer Science,~~\\ Nanyang Technological University\\
Singapore\\
lvpeizhuo@gmail.com\\
~\\
~}
\and
\IEEEauthorblockN{Yan Wang}
\IEEEauthorblockA{Institute of Information Engineering,\\ Chinese Academy of Sciences\\
Beijing University of Technology\\
wangyan@iie.ac.cn\\
~}
\and
\IEEEauthorblockN{Rujie Dai}
\IEEEauthorblockA{Institute of Information Engineering,\\ Chinese Academy of Sciences\\
dairujie2024@iie.ac.cn\\
~\\
~}
\and
\IEEEauthorblockN{Guokun Xu}
\IEEEauthorblockA{Institute of Information Engineering,\\ Chinese Academy of Sciences\\
xuguokun@iie.ac.cn\\
~\\
~}
\and
\IEEEauthorblockN{Qiujian Lv}
\IEEEauthorblockA{Institute of Information Engineering,\\ Chinese Academy of Sciences\\
lvqiujian@iie.ac.cn\\
~\\
~}
\and
\IEEEauthorblockN{Hangcheng Liu}
\IEEEauthorblockA{Nanyang Technological University\\
hangcheng.liu@ntu.edu.sg\\
~\\
~}
\and
\IEEEauthorblockN{Weiqing Huang}
\IEEEauthorblockA{Institute of Information Engineering,\\ Chinese Academy of Sciences\\
huangweiqing@iie.ac.cn\\
~\\
~}
\and
\IEEEauthorblockN{Wei Dong}
\IEEEauthorblockA{Nanyang Technological University\\
wei\_dong@ntu.edu.sg\\
~\\
~}
\and
\IEEEauthorblockN{Jiaheng Zhang}
\IEEEauthorblockA{National University of Singapore\\
jhzhang@nus.edu.sg\\
~\\
~}
}
\author{
\IEEEauthorblockN{
Weijie Wang\IEEEauthorrefmark{1,}\IEEEauthorrefmark{2,}\textsuperscript{*},
Peizhuo Lv\IEEEauthorrefmark{3,}\textsuperscript{*,}\textsuperscript{\dag},
Yan Wang\IEEEauthorrefmark{1,}\IEEEauthorrefmark{5,}\textsuperscript{\dag},
Rujie Dai\IEEEauthorrefmark{1},
Guokun Xu\IEEEauthorrefmark{1},
Qiujian Lv\IEEEauthorrefmark{1},
Hangcheng Liu\IEEEauthorrefmark{3},\\
Weiqing Huang\IEEEauthorrefmark{1},
Wei Dong\IEEEauthorrefmark{3},
Jiaheng Zhang\IEEEauthorrefmark{2}}
\IEEEauthorblockA{\IEEEauthorrefmark{1}Institute of Information Engineering, Chinese Academy of Sciences}
\IEEEauthorblockA{\IEEEauthorrefmark{2}National University of Singapore}
\IEEEauthorblockA{\IEEEauthorrefmark{3}Nanyang Technological University}
\IEEEauthorblockA{\IEEEauthorrefmark{4}Beijing University of Technology}
\thanks{\textsuperscript{*} Co-first authors. \textsuperscript{\dag} Co-corresponding authors.}

Email: wangweijie@iie.ac.cn, lvpeizhuo@gmail.com, \{wangyan,dairujie2024,xuguokun,lvqiujian\}@iie.ac.cn,\\ hangcheng.liu@ntu.edu.sg, huangweiqing@iie.ac.cn, wei\_dong@ntu.edu.sg, jhzhang@nus.edu.sg
}
\IEEEoverridecommandlockouts
\makeatletter\def\@IEEEpubidpullup{1.5\baselineskip}\makeatother
\IEEEpubid{\parbox{\columnwidth}{
		The first author conducted this work as a visiting scholar at National University of Singapore, whose support is gratefully acknowledged.
}
\hspace{\columnsep}\makebox[\columnwidth]{}}

\maketitle

\begin{abstract}
Graph Retrieval-Augmented Generation (GraphRAG) has emerged as a key technique for enhancing Large Language Models (LLMs) with proprietary Knowledge Graphs (KGs) in knowledge-intensive applications. As these KGs often represent an organization's highly valuable intellectual property (IP), they face a significant risk of theft for private use. In this scenario, attackers operate in isolated environments. This private-use threat renders passive defenses like watermarking ineffective, as they require output access for detection. Simultaneously, the low-latency demands of GraphRAG make strong encryption which incurs prohibitive overhead impractical. To address these challenges, we propose \myname, a novel framework based on Data Adulteration designed to make any stolen KG unusable to an adversary. Our framework pre-emptively injects plausible but false adulterants into the KG. For an attacker, these adulterants deteriorate the retrieved context and lead to factually incorrect responses. Conversely, for authorized users, a secret key enables the efficient filtering of all adulterants via encrypted metadata tags before they are passed to the LLM, ensuring query results remain completely accurate. Our evaluation demonstrates the effectiveness of this approach: \myname degrades the performance of unauthorized systems to an accuracy of just 5.3\%, while maintaining 100\% fidelity for authorized users with negligible overhead. Furthermore, \myname proves robust against various sanitization attempts, retaining 80.2\% of its adulterants.
\end{abstract}

\IEEEpeerreviewmaketitle

\section{Introduction}
\label{sec:introduction}

Graph Retrieval-Augmented Generation (GraphRAG) has been widely applied to mitigate the hallucination in Large Language Models (LLMs) and enhance their capabilities in domain-specific, knowledge-intensive, and privacy-sensitive tasks. For example, Pfizer~\cite{su2024Pfizer} utilizes it to accelerate drug discovery, and Siemens employs it to power its smart manufacturing strategy~\cite{hubauer2018siemens}. The proprietary Knowledge Graphs (KGs) that power these advanced systems store domain-specific knowledge, including vast amounts of intellectual property and sensitive information. Although the construction costs for these commercial KGs are not publicly disclosed, the required investment is substantial. This scale is illustrated by expert-driven projects like Cyc~\cite{lenat1995cyc}, which required an estimated \$120 million, translating to roughly \$5.71 per factual statement~\cite{paulheim2018much}.

This high value makes proprietary KGs a prime target for IP theft. An attacker might steal the KG through external cyber intrusions or by leveraging malicious insiders. For instance, a star engineer stole over 14,000 proprietary files for Waymo's LiDAR system and then joined a competitor~\cite{Chen2018WaymoUber} and the 2020 hack of the European Medicines Agency to obtain Pfizer/BioNTech's confidential vaccine data~\cite{national2020advisory}. Once an attacker has successfully stolen a KG, they can deploy it in a private GraphRAG system to replicate the powerful capabilities, avoiding costly investments. The gravity of this threat is recognized in major regulatory frameworks, both the EU AI Act~\cite{madiega2021aiact} and the NIST AI Risk Management Framework~\cite{ai2023nist} emphasize the need for robust data security and resilience, emphasizing the importance of developing effective KG protection.

Protecting the KGs within GraphRAG systems is particularly challenging due to the threat of private use and the performance requirements of GraphRAG. The private deployment scenario, where an attacker operates the stolen KG in an isolated environment, makes common defenses like digital watermarking~\cite{lv2025rag} ineffective. Watermarking requires access to a system's output to trace leaks. However, the owner cannot get the outputs in a private setting, so it is impossible to detect misuse or hold the attacker accountable. Simultaneously, the low-latency requirements of interactive GraphRAG make strong cryptographic solutions (e.g., homomorphic encryption) impractical. Fully encrypting the text and embeddings would require decrypting large portions of the graph for every query. This process introduces prohibitive computational overhead and latency, making it unsuitable for real-world use~\cite{zhou2025privacy}.

To address these challenges, we propose \myname (Active Utility Reduction via Adulteration), a novel framework that makes a stolen KG unusable to an adversary while maintaining minimal performance overhead for the GraphRAG system. \myname achieves this by utilizing data adulteration to inject plausible but false information into the KG. To ensure both efficiency and robustness, the process begins by identifying a minimal set of critical nodes for maximum impact. It then employs a hybrid generation strategy to create sophisticated adulterants that are plausible at both the semantic and structural levels. Because our goal is to render the KG unusable in unauthorized GraphRAG systems, we select from the generated adulterants only those with the most significant impact on the LLM output for injection. Consequently, when an attacker uses a stolen copy, these adulterants will be retrieved as context, deteriorating the LLM's reasoning and leading to factually incorrect responses. Conversely, the authorized system uses a secret key only known to the owner to identify and filter all adulterants via their encrypted metadata tags before they are passed to the LLM, ensuring query results remain completely accurate. This design ensures that only a user possessing the secret key can distinguish the injected adulterants from the authentic data, fundamentally coupling the KG's utility to the key's secrecy and against private use.

We evaluate \myname on four benchmark datasets (MetaQA, WebQSP, FB15k-237, and HotpotQA) and across four different LLMs (GPT-4o, Gemini-2.5-flash, Qwen-2.5-7B, and Llama2-7B). First, \myname proves highly effective at degrading the utility of a stolen KG for unauthorized users, degrades the performance of unauthorized systems to an accuracy of just 4.4\% to 5.3\%. Second, our method maintains perfect fidelity for authorized users, achieving 100\% performance alignment with the original clean system while incurring minimal overhead, a maximum query latency increase under 14\%. Third, \myname is highly stealthy; its adulterants evade both structural and semantic anomaly detectors, with detection rates below 4.1\%. Fourth, the defense is robust; even after facing advanced sanitization attacks, the unauthorized system's accuracy remains as low as 17.7\%. Finally, extensive evaluations across various system parameters confirm the stability and effectiveness of our approach.

Our main contributions are summarized as follows:
\begin{itemize}

    \item We are the first to identify and address the security threat of private use for stolen proprietary KG. In this scenario, traditional IP protection methods like watermarking and encryption are ineffective. 
    \item We introduce a novel defense paradigm for knowledge graphs that shifts the focus from passive detection to active degradation of the value of stolen assets, providing a fundamental solution to IP protection, especially against the threat of private use. We realize this paradigm with \myname, a framework that automates the generation and injection of stealthy, robust, and filterable adulterants.
    \item  We conduct a comprehensive evaluation across four datasets and four LLMs, and the results demonstrate that \myname meets all defense requirements, achieving high effectiveness, perfect fidelity, and strong robustness.
\end{itemize}

\section{Background}
\label{sec:background}

\subsection{Graph Retrieval-Augmented Generation}
Retrieval-augmented generation (RAG) enhances Large Language Models (LLMs) by grounding them in external knowledge, which reduces hallucinations and provides timely information~\cite{lewis2020retrieval, gao2023retrieval, fan2024survey}. However, conventional RAG operating on unstructured text struggles to capture complex relationships, leading to incomplete context and reasoning failures~\cite{zhang2024found}. GraphRAG addresses this by leveraging a Knowledge Graph (KG) as its external knowledge source~\cite{peng2024graph, edge2024local, he2024g}.

Formally, the goal of GraphRAG is to find the optimal answer $a^*$ for a given query $q$ and a graph $G$, which can be defined as:

\begin{equation}
a^* = \arg\max_{a \in A} p(a|q, G)
\end{equation}

Where $A$ is the set of all possible responses. This target distribution is then jointly modeled with a graph retriever $p_{\theta}(G'|q, G)$ and an answer generator $p_{\phi}(a|q, G')$, where $\theta$ and $\phi$ are learnable parameters. The process is decomposed as:

\begin{equation}
\begin{split}
p(a|q, G) & = \sum_{G' \subseteq G} p_{\phi}(a|q, G') p_{\theta}(G'|q, G) \\
          & \approx p_{\phi}(a|q, G^*)
\end{split}
\end{equation}

Where $G^*$ is the optimal subgraph retrieved from the full graph $G$, because the number of candidate subgraphs can grow exponentially, the process is approximated by first employing a retriever to extract the most relevant subgraph $G^*$, after which the generator produces the answer based on this retrieved subgraph.

By operating directly on graph structures, GraphRAG can explicitly model and traverse relationships between entities, enabling more precise retrieval of paths~\cite{luo2024reasoning, sun2024thinkongraph} or subgraphs, and supporting complex reasoning tasks far more effectively than text-based RAG systems~\cite{jiang2023structgpt, jin2024graph}. The retrieved graph structures can then be serialized into various formats, such as adjacency lists~\cite{guo2023gpt4graph} or text sequences~\cite{zhao2023graphtext}, for the LLM to process. 

The practical value and potential of GraphRAG are underscored by its widespread adoption across the technology industry. Major corporations are actively developing and deploying GraphRAG solutions to leverage their vast data ecosystems. For instance, Microsoft has open-sourced its Project GraphRAG for complex data exploration~\cite{larson2024microsoft}. Google provides reference architectures for building KGQA systems on its cloud platform~\cite{google2024vertexai}. Meanwhile, Alibaba focuses on high-performance GNN-based retrieval for e-commerce~\cite{2025alibaba}. This broad industrial investment underscores the critical importance of KGs as core assets in modern RAG systems, highlighting the urgent need for robust protection methods.

\subsection{Link Prediction Model}
Link prediction is a fundamental task for knowledge graph completion, which aims to infer missing links (or triples) based on existing ones~\cite{wang2021survey}. This process is essential for augmenting the coverage and utility of knowledge graphs, which are frequently incomplete due to the vastness and dynamic nature of real-world information. The dominant approach for this task is Knowledge Graph Embedding (KGE), where entities and relations are mapped to a low-dimensional continuous vector space. Within this embedding space, a scoring function $f(h, r, t)$ is employed to quantify the plausibility of a candidate triple $(h, r, t)$, thus enabling the identification of likely but unobserved facts.

KGE models can be broadly categorized according to the structure of their scoring functions~\cite{wang2021survey}. Early approaches, such as translational distance models including TransE~\cite{bordes2013translating} and RotatE~\cite{sun2019rotate}, represent relations as operations that translate or rotate entity embeddings in the vector space. These models offer efficient and interpretable mechanisms for capturing relational patterns. In contrast, more recent advancements leverage neural architectures, exemplified by models like ConvE~\cite{dettmers2018convolutional}, which employ convolutional neural networks to learn complex, non-linear interactions between entities and relations. 


In this paper, we leverage link prediction models not for their traditional purpose of discovering new knowledge, but as a core component of \myname to generate adulterants.

\section{Problem Statement}
\label{sec:problem}
\begin{figure}[t]
    \centering
    \includegraphics[width=0.98\linewidth]{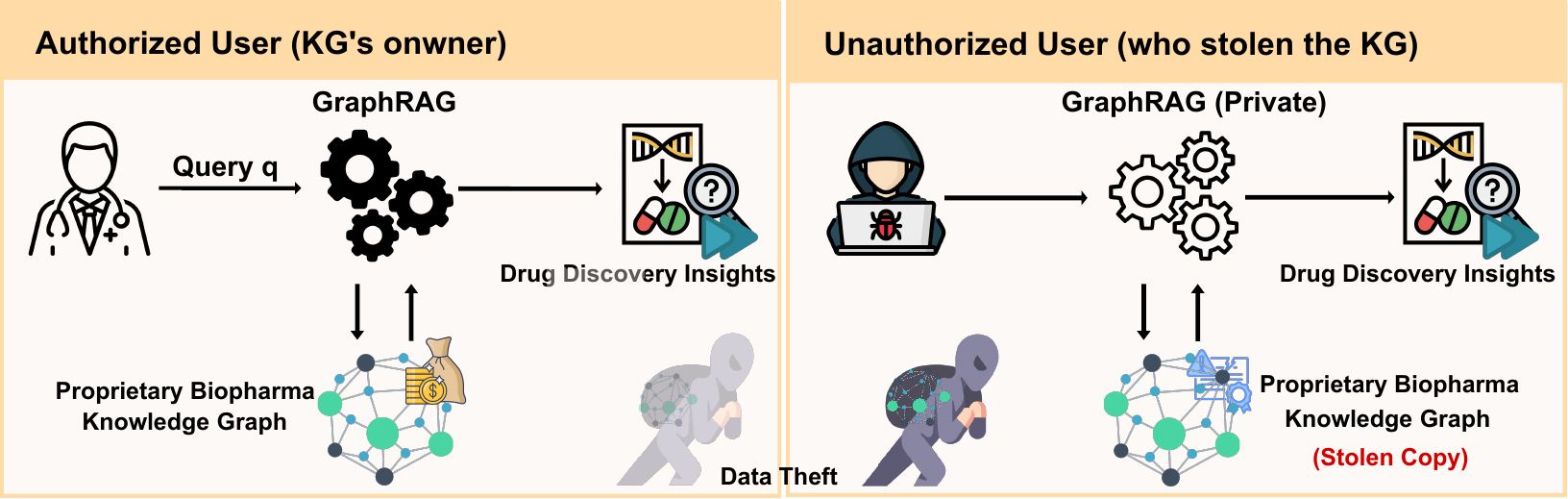}
    \caption{By stealing the Knowledge Graph of a GraphRAG system, the adversary can almost completely replicate the original system's functionality in a private environment avoiding costly investments.}
    \label{fig:motivating_example}
\end{figure}

As proprietary KGs become increasingly central to high-value applications, they emerge as prime targets for IP theft. This section defines the threat model we address and outlines the core requirements for an effective defense.

\subsection{An Motivating Example}

Consider a high-value, proprietary KG developed by a biopharmaceutical firm (e.g., Pfizer). This KG, representing complex interactions between compounds, genes, and diseases, is the core asset behind their commercial GraphRAG service, which helps researchers accelerate the pace of drug discovery. The immense value provided by this KG also makes it a prime target for theft. As the examples in our introduction demonstrate, these threats are tangible and include malicious insiders exfiltrating data and external actors breaching defenses via targeted cyberattacks.

After successfully stealing the KG, the attacker integrates the asset into their own privately hosted GraphRAG system, as shown in Figure~\ref{fig:motivating_example}. This allows them to accelerate their internal research and development by exploiting the owner's curated knowledge to gain analytical insights, thereby avoiding costly investments. This scenario is realistic because the attacker lacks the specialized expertise and resources to build or validate such a complex KG from scratch, which is the primary motivation for the theft. If the stolen data is directly usable, the owner would suffer substantial losses. This highlights the urgent need for a defense that persists within the data itself, preventing the unauthorized use of the KG post-theft.

\subsection{Threat Model}

Our threat model considers the post-theft scenario illustrated in Figure~\ref{fig:motivating_example}, where an attacker, having stolen a copy of a proprietary KG, aims to integrate it into their own GraphRAG system to exploit it for their benefit.

\subsubsection{Attacker's Target}
The attacker's target is to steal the proprietary KG to deploy their own private GraphRAG system. By doing so, they aim to replicate the core functionalities of the owner's service and profit from it.

\subsubsection{Attacker's Capabilities}
We model a sophisticated adversary with the following capabilities:

\paragraph{Independent System Deployment} 
The attacker can build and operate their own GraphRAG system in a private, offline environment. The attacker may therefore use any retriever models and Large Language Models (LLMs) they want. Consequently, any defense mechanisms embedded within the owner's original GraphRAG components (e.g., backdoors in the retriever or LLM) are ineffective. 

\paragraph{Knowledge Sanitization Attempts}
We assume the attacker is not naive and may suspect that the KG is protected. For high-value KGs, such as Pfizer's drug discovery graph, conducting a thorough validation and sanitization of the KG would require a team of domain experts. However, an adversary possessing such a team would likely have the capability to build their own KG, negating the original incentive for theft. This paradox makes any such expert-driven attempt to sanitize the data expensive and strategically illogical. So the attacker may employ standard sanitization methods from the knowledge graph domain, such as structural or semantic detection.

\paragraph{Lack of Secret Key}
As is commonly assumed in prior research~\cite{zhou2025privacy,qu2024provably,chen2024eff}, we assume the secret key is secure. The attacker cannot leverage it to identify and filter out the adulterants.

\subsection{The Philosophy of Our Approach}
\begin{figure*}[t]
    \centering
    \includegraphics[width=0.95\linewidth]{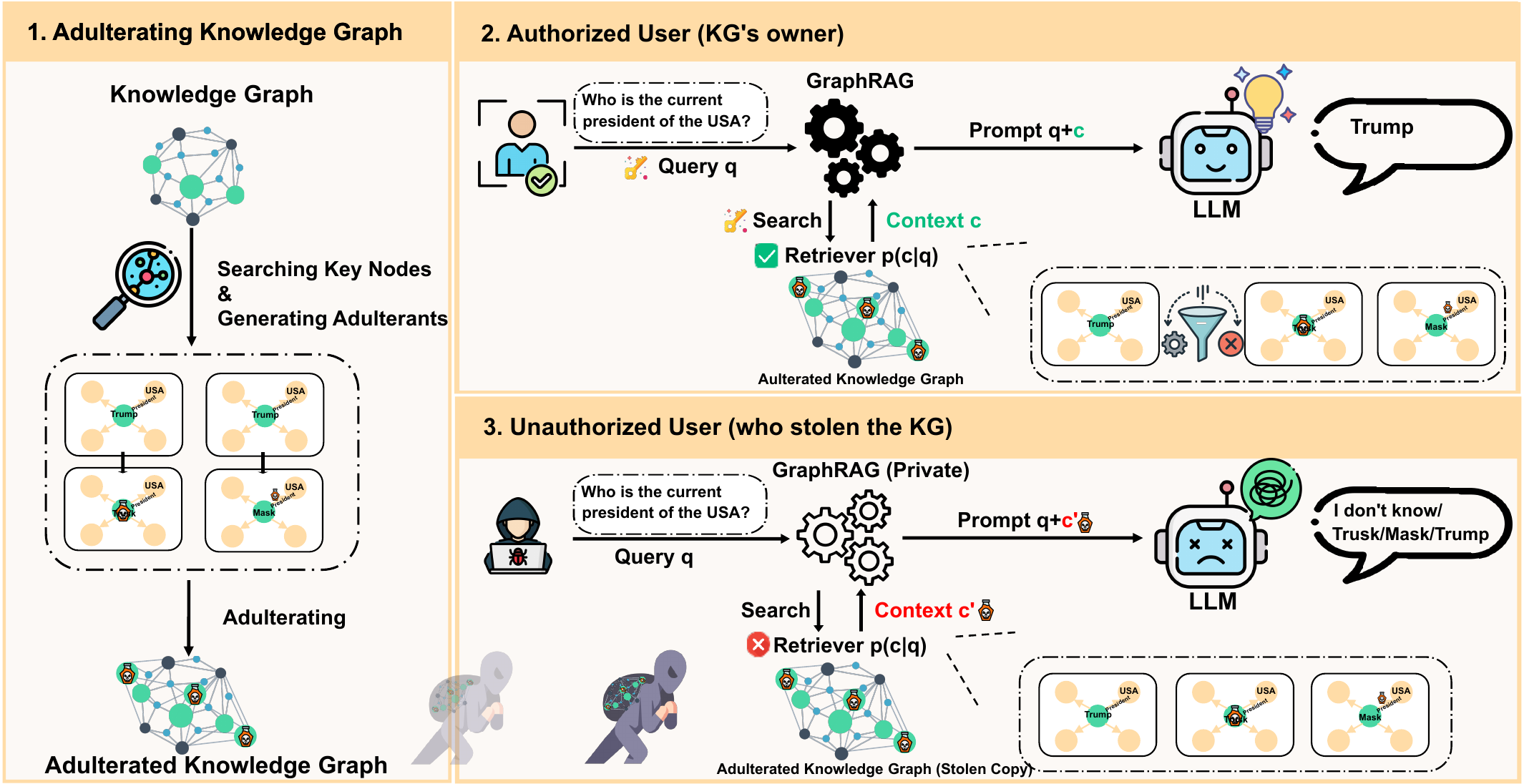}
    \caption{By adulterating the KG, we create an information asymmetry. The authorized user can filter the adulterants to get the correct answer, while the unauthorized user's LLM is misled by the adulterated context.}
  
    \label{fig:preliminary_insight}
\end{figure*}

We leverage the operational mechanism of GraphRAG systems to achieve IP protection. In a typical workflow, when a user submits a query $q$, the retriever typically identifies a target node within the $KG$ and then retrieves a relevant subgraph, consisting of neighboring nodes and their corresponding relationships (edges), as the context $c$. The final answer is then produced by the LLM conditioned on this context. Since an attacker's goal is to replicate the owner's service by using the stolen KG to retrieve useful context for their own LLM. If the retrieval process is corrupted such that the retrieved context becomes misleading or harmful, the value of the stolen $KG$ is significantly degraded, thereby preventing unauthorized use.

Building on this insight, we propose injecting a set of false but plausible adulterants $A$, into the original $KG$ to create an adulterated version $KG' = KG \cup A$. As shown in Figure~\ref{fig:preliminary_insight}, this creates an information asymmetry: for an unauthorized user, the retrieval on the stolen $KG'$ gets an adulterated context $c' = R(q, KG')$, which contains both authentic facts and adulterants ($c' = c \cup A$). Feeding this context into the LLM will lead to factually incorrect or nonsensical responses, making the KG useless. However, such a mechanism would also affect authorized users, as their GraphRAG system also operates on the adulterated $KG'$. Therefore, a filtering mechanism is required. For an authorized user who possesses a secret key $k$, the system employs a filtering mechanism $F$. After retrieving the contaminated context $c'$, the filter purifies it: $c_{clean} = F(c', k)$, ensuring that the adulterants are removed. The LLM thus processes only the correct context, preserving the system's utility and accuracy.

Instead of focusing on preventing unauthorized data access, our approach focuses on the data's usability. The security of \myname is thereby coupled to the secrecy of the key. Even if an attacker were to discover and replicate our filtering mechanism, it would be entirely ineffective without the correct key.

\subsection{Defending Requirements}

To achieve this, an effective defense mechanism must satisfy the following core requirements:

\begin{itemize}
    \item \textbf{Effectiveness:} The effectiveness of \myname is measured by the degree of performance degradation inflicted upon an unauthorized GraphRAG system. Specifically, the injected adulterants must cause a significant drop in answer accuracy and a rise in content hallucinations. 
    \item \textbf{Fidelity:} The defense should not significantly impact authorized users. The system's performance, in terms of both retrieval and final answer quality, must remain nearly identical to that of the original system. 
    \item \textbf{Stealthiness:} It should be difficult for attackers to discover or detect adulterants from the stolen KG.
    \item \textbf{Robustness:} The adulterants should remain effective even if the attacker attempts to cleanse the KG or employs different GraphRAG architectures (e.g., various retrievers and LLMs). 
\end{itemize}

\section{Approach}
\label{sec:method}
\begin{figure*}[t]
    \centering
    \includegraphics[width=0.98\linewidth]{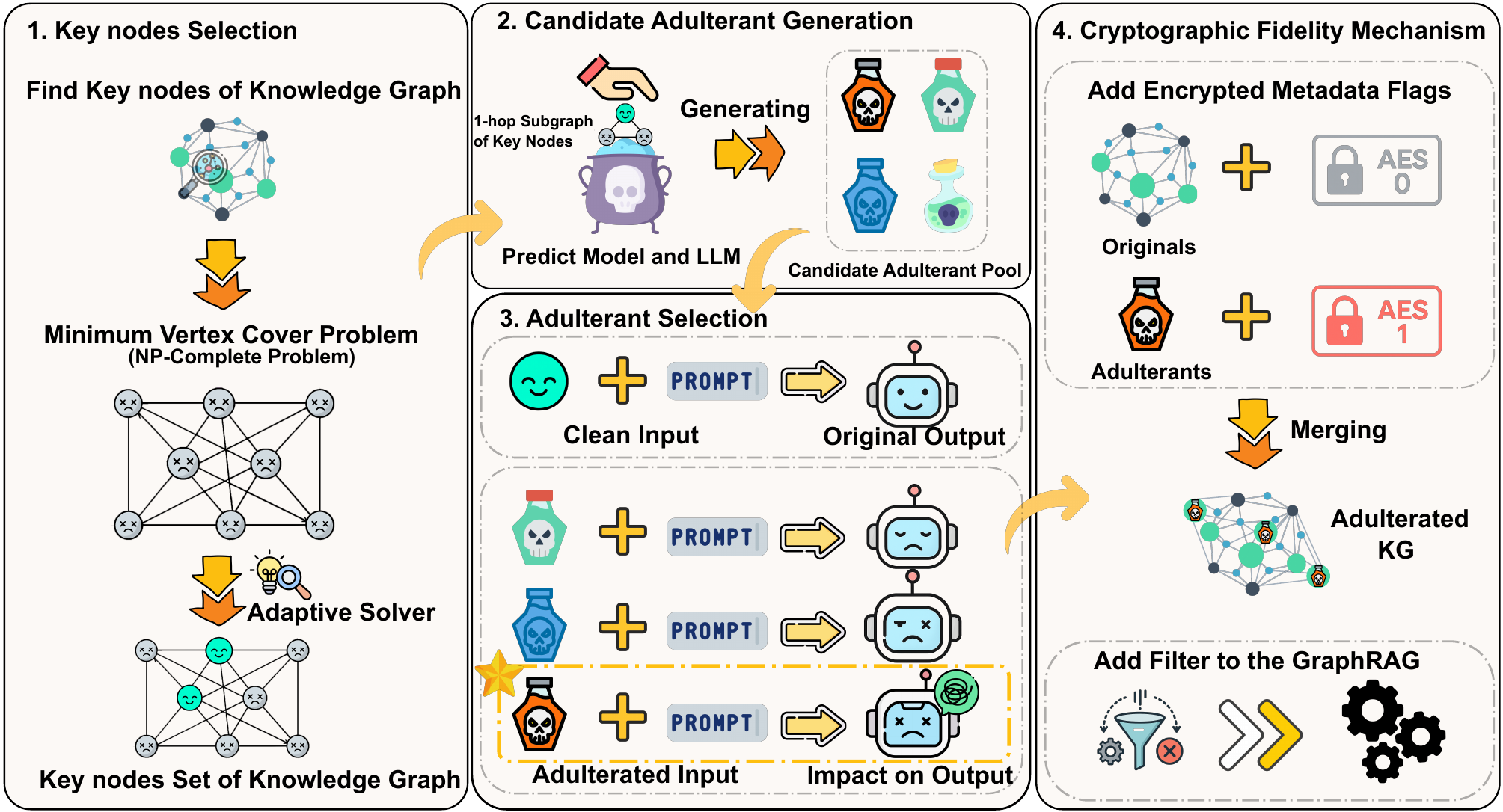}
    \caption{Overview of the \myname}

    \label{fig:framework_overview}
\end{figure*}
\subsection{Overview}

The \myname\ is an automated pipeline that transforms a standard Knowledge Graph (KG) into a self-defending asset through a four-stage process, as illustrated in Figure~\ref{fig:framework_overview}. To ensure the adulterants can efficiently impact every relationship in the KG, we map the target selection process to the Minimum Vertex Cover (MVC) problem and solve this NP-complete problem with a specialized hybrid algorithm. Considering the inherent limitations of any single generation method, we then employ a hybrid strategy: link prediction models create structurally plausible but false edges. In contrast, LLMs generate semantically coherent but fake nodes, resulting in candidates that are plausible on both structural and semantic levels. Recognizing that plausibility does not guarantee effectiveness, the third stage performs impact-driven selection, utilizing the Semantic Deviation Score (SDS) to optimize for the adulterants' destructive ability on the final LLM output. Finally, we establish a cryptographic fidelity mechanism that provides a provably secure guarantee for authorized users to remove all adulterants, thereby achieving high-fidelity performance.

In the following subsections, we provide a detailed explanation of each of these stages.

\subsection{MVC-based Key Nodes Selection}
This initial stage of our framework confronts a fundamental trade-off in graph-based defense: achieving maximum defensive impact with minimal costs. While generating adulterants is necessary for defense, it inevitably expands the graph, increasing storage overhead and query latency. The core challenge, therefore, is to identify the smallest set of nodes for adulteration that can still guarantee graph-wide influence. Successfully identifying this set not only enhances stealth by minimizing the number of modifications but also reduces the overall defense cost, which includes both the computational expense of generation and the overhead from the resulting graph expansion. This stage is thus designed to find an optimal solution to this complex optimization problem.

A common approach for this task is to employ graph-theoretic importance metrics, such as degree centrality or PageRank. However, these methods are insufficient for our purposes as they fail to guarantee comprehensive coverage; they may identify a cluster of influential nodes but leave entire subgraphs untouched. To ensure that our defense can influence \textit{every} relationship in the KG, we formally redefine the selection task. Thus, the objective shifts from identifying the most influential nodes to determining the smallest set of nodes adjacent to every edge in the graph. Let $V_{\text{adulterant}}$ be the set of nodes selected for adulteration. Our objective is to find a set $V_{\text{adulterant}} \subseteq V$ that minimizes its size, $|V_{\text{adulterant}}|$, subject to the constraint that for every edge $(u, v) \in E$, the condition $\{u, v\} \cap V_{\text{adulterant}} \neq \emptyset$ must hold. This formulation is precisely the definition of the Minimum Vertex Cover (MVC) problem, a classic NP-complete problem in graph theory.

Formally, given a graph $G = (V, E)$, a vertex cover is a subset of vertices $C \subseteq V$ such that for every edge $(u, v) \in E$, at least one of its endpoints is in the cover (i.e., $u \in C$ or $v \in C$). The MVC problem seeks to find a vertex cover $C^*$ with the minimum possible cardinality. This is expressed as the following optimization problem:

\begin{equation}
\label{eq:mvc}
C^* = \arg\min_{C \subseteq V} |C| \quad \text{s.t.} \quad \forall (u, v) \in E, u \in C \lor v \in C.
\end{equation}

The NP-complete nature of this problem introduces the computational challenge of finding a perfectly optimal solution. The state-of-the-art approach for determining the exact solution is to use an Integer Linear Programming (ILP) solver. We formulate the MVC problem as an ILP by introducing a binary variable $x_v$ for each vertex $v \in V$, where $x_v=1$ if the vertex is selected for the cover, and $x_v=0$ otherwise. The problem is then expressed as:

\begin{equation}
\label{eq:ilp}
\begin{aligned}
& \underset{x}{\text{minimize}}
& & \sum_{v \in V} x_v \\
& \text{subject to}
& & x_u + x_v \ge 1, \quad \forall (u, v) \in E \\
& & & x_v \in \{0, 1\}, \quad \forall v \in V.
\end{aligned}
\end{equation}

A solver (e.g., Pulp~\cite{mitchell2011pulp}, Gurobi~\cite{pedroso2011gurobi}) can process this formulation to guarantee finding the globally optimal vertex cover. However, the primary drawback of ILP solvers is their computational complexity, which grows exponentially with the size of the problem. This makes them computationally prohibitive for large-scale graphs.

Through empirical analysis, we identified a practical boundary for the feasibility of the ILP-based approach. We observed that for graphs exceeding approximately 150,000 nodes, the runtime of the exact solver becomes impractical for real-world applications. A detailed analysis of this trade-off between computation time and solution optimality is presented in Appendix~\ref{sec:appendix_threshold}.

To balance solution optimality with computational feasibility, we employ an adaptive strategy. For small to medium-scale graphs (up to 150,000 nodes), we utilize the ILP solver to obtain the guaranteed optimal key node set. For larger graphs, we switch to a heuristic based on the Malatya algorithm~\cite{karci2022malatya}. This polynomial-time algorithm consists of a two-step iterative process. First, it calculates a Malatya centrality value for each node $u$ in the current graph, which is defined as:

\begin{equation}
\label{eq:malatya}
MC(u) = \sum_{v \in \mathcal{N}(u)} \frac{\text{deg}(u)}{\text{deg}(v)}
\end{equation}

Where $\mathcal{N}(u)$ is the set of neighbors of node $u$, and $\text{deg}(\cdot)$ is the degree of a node. Second, it selects the node with the maximum centrality, adds it to the solution set, and then removes the node and its incident edges from the graph. This process repeats until no edges remain, providing a high-performance, scalable approach to finding a near-optimal vertex cover. The entire adaptive process is detailed in Appendix~\ref{sec:appendix_algorism}.

\subsection{Hybrid Adulterant Generation}
\label{sec:candidate_generation}
Upon identifying the key nodes $V_{\text{adulterant}}$, the next critical step is to generate a pool of candidate adulterants. The primary objective of this stage is to achieve maximum stealthiness; the generated adulterants must be indistinguishable from legitimate data to evade detection. The core challenge lies in the dual nature of plausibility within a knowledge graph: an effective adulterant must possess both structural plausibility and semantic coherence. A structurally plausible adulterant conforms to the existing graph topology. At the same time, a semantically coherent one aligns with the real-world meaning of the entities it connects. Relying on a single generation technique often fails to satisfy both requirements simultaneously. To overcome this, we propose a hybrid generation method. This approach combines the strengths of two distinct methodologies: we leverage link prediction models, which excel at maintaining structural integrity, and LLMs, which are good at generating semantically meaningful content. By combining these, we create a diverse set of candidate adulterants designed to deteriorate different retrieval scenarios, which we detail in the following subsections. Let the KG be $G = (V, \mathcal{R}, E)$, where $V$ is the set of entities, $\mathcal{R}$ is the set of relation types, and $E$ is the set of triples $(h, r, t)$ with $h, t \in V$ and $r \in \mathcal{R}$. 

\subsubsection{Generating Adulterated Edges via Link Prediction}
This method targets the retrieval scenario where a query targets a key node to retrieve its incident edges and adjacent nodes. For this scenario, we generate structurally plausible but factually incorrect edges. The objective is to create adulterated relationships that match the KG's existing topology. For this purpose, we leverage a pre-trained link prediction model, $f_{LP}$. Such models are trained to score the likelihood of a given triple $(h, r, t)$ being true, making them appropriate for identifying high-plausibility but non-existent relationships (edges) connected to the key nodes.

Specifically, for each key node $v_k \in V_{\text{adulterant}}$, we iterate over all true triples $(h, r, t) \in E$ where $v_k$ is either the head or the tail entity. If $v_k$ is the head, we query the model with $(v_k, r, ?)$, which returns a ranked list of candidate tail entities $L_{t'} = [t'_1, t'_2, \dots]$. To ensure the generated triples are factually incorrect, we form a set of candidate adulterated triples, $P_{\text{edge}}(v_k, r, t)$, by selecting the top-$N$ entities from $L_{t'}$ that are not the ground-truth tail $t$:
\begin{equation}
P_{\text{edge}}(v_k, r, t) = \{ (v_k, r, t'_i) \mid t'_i \in \text{top-}N(L_{t'}) \land t'_i \neq t \}
\end{equation}
$N$ is a tunable hyperparameter that balances the potency and stealth. A similar process is applied when $v_k$ is the tail. The set of all candidate adulterated edges, denoted $C_{\text{edge}}$, is formally defined as the union of all such adulterant sets generated from all key nodes and their incident relations:
\begin{equation}
\begin{split}
C_{\text{edge}} = & \{ p \mid \exists v_k \in V_{\text{adulterant}}, \exists (h,r,t) \in E \\ 
& \text{ s.t. } (h=v_k \lor t=v_k) \land p \in P_{\text{edge}}(h,r,t) \}
\end{split}
\end{equation}

\subsubsection{Generating Adulterated Nodes via LLM}
This method targets a distinct retrieval scenario: queries originating from the neighbors of a key node, where the key node itself is the intended result. For this scenario, we generate entirely new, semantically coherent but factually incorrect nodes. The objective is to create an adulterated node that is semantically similar to the original key node, causing it to appear with the real one in search results. For this purpose, we leverage a LLM, $f_{LLM}$. Such models excel at understanding context and generating coherent text, making them perfectly suited for creating plausible but fake entity names. The prompt template used for this generation is detailed in Appendix~\ref{prompt_generateadulterant}.

To make this adulterated node structurally convincing, we replicate the entire local neighborhood of the original key node, creating a new set of adulterated triples $P_{\text{node}}(v_{k})$ where $v'_k$ replaces $v_k$. The corresponding set of adulterated triples is created as follows:

\begin{equation}
\begin{split}
P_{\text{node}}(v'_{k}) & = \{ (v'_{k}, r, t) \mid (v_k, r, t) \in E \} \\
& \cup \{ (h, r, v'_{k}) \mid (h, r, v_k) \in E \}
\end{split}
\end{equation}

The complete set of adulterated triples generated via this node-creation strategy, denoted $C_{\text{node}}$, is the union of adulterant sets from all adulterated entities:
\begin{equation}
C_{\text{node}} = \{ p \mid \exists v_k \in V_{\text{adulterant}} \text{ s.t. } p \in P_{\text{node}}(v'_{k}) \}
\end{equation}

\subsubsection{Final Candidate Pool}
Finally, the outputs of these two methods are aggregated to form the final candidate adulterant set, $C$. This set will proceed to the subsequent selection stage to identify the optimal subset of adulterants for injection.
\begin{equation}
\label{eq:candidate_set}
C = C_{\text{edge}} \cup C_{\text{node}}
\end{equation}

\subsection{Impact-Driven adulterant Selection}
Following the generation of a diverse candidate pool, the next stage is to select the most effective adulterants for injection. The plausibility of an adulterant does not guarantee its deteriorating impact on the final output of an adversary's LLM. We introduce an impact-driven selection process that directly measures each candidate's ability to semantically alter the answers generated by a GraphRAG system.

The selection process is as follows: for each key node $v_k \in V_{\text{adulterant}}$, we first assemble its local candidate set, $C(v_k)$, which includes all adulterated edges and adulterated nodes generated specifically for $v_k$. We then evaluate each candidate adulterant $c \in C(v_k)$ by simulating its effect on a GraphRAG system's output.

To quantify this deteriorative potential, we propose the Semantic Deviation Score (SDS). This metric is designed to measure the semantic shift in the system's output caused by the presence of a single candidate adulterant. We define the SDS as the Euclidean distance between the sentence embeddings of the answers generated with and without the adulterant. This metric directly aligns with our objective of deteriorating the final output; a greater semantic distance signifies a more significant and thus more effective attack. By leveraging pre-trained sentence embedding models, we can capture nuanced semantic differences that simple lexical comparisons would miss. Formally, the SDS for a candidate adulterant $c$ concerning a question $q$ is defined as:

\begin{equation}
\label{eq:sds}
SDS(c, q) =  ||\mathcal{E}(\mathcal{S}(q, G)) - \mathcal{E}(\mathcal{S}(q, G \cup \{c\}))||_2
\end{equation}

Where $\mathcal{E}(\cdot)$ is a sentence embedding function, $\mathcal{S}(q, G)$ is the answer generated for question $q$ using the original graph $G$, and $G \cup \{c\}$ represents the graph with the candidate adulterant injected.

For each key node $v_k \in V_{\text{adulterant}}$, we first assemble its local candidate set, $C(v_k)$, which includes all adulterated edges and nodes generated for $v_k$. We then evaluate each candidate adulterant $c \in C(v_k)$ by simulating its effect on a GraphRAG system's output. To obtain a robust score, we use a set of questions $Q$ sourced from an open-source Question-Answering (QA) dataset relevant to the KG's domain. LLM can also generate this $Q$ detailed in Appendix~\ref{prompt_generateQA}. The final score for a candidate is the average SDS over all questions. The selection process is detailed in Appendix~\ref{sec:appendix_algorism}.

After calculating the SDS for all candidates in $C(v_k)$, we select the one with the highest score as the final adulterant for that key node:
\begin{equation}
\label{eq:select_best}
p^*_k = \arg\max_{c \in C(v_k)} SDS(c)
\end{equation}

This process is repeated for every key node in $V_{\text{adulterant}}$. The final set of adulterants to be injected into the KG, denoted $P^*$, is the union of the most impactful adulterants selected for each key node:
\begin{equation}
\label{eq:final_adulterants}
P^* = \{p^*_k \mid v_k \in V_{\text{adulterant}}\}
\end{equation}

\subsection{Cryptographic Fidelity Mechanism}
A fundamental requirement for any practical defense system is to guarantee data fidelity for authorized users. While the injected adulterants are designed to mislead adversaries, they must be removable for an authorized system. The fundamental difficulty lies in designing a filtering mechanism that provides discriminability for the authorized system while being indistinguishable from an adversary. To overcome this, we introduce the cryptographic fidelity mechanism. This approach moves beyond fallible heuristics by embedding an encrypted metadata flag into every node and edge. By leveraging symmetric-key encryption, we provide a provably secure guarantee that only users possessing the secret key $K_{\text{owner}}$ can decrypt these flags and perfectly reconstruct the original, untainted graph, thereby achieving high-fidelity performance.

When populating the graph database, we embed a special metadata property into every node and edge in the final graph $G'=(V', E')$. To ensure stealthiness, this property is given a common name, such as `remark' or `annotation', making it appear as a standard, non-critical attribute. The value of this property, a ciphertext, is generated for each node and edge using the owner's secret key $K_{\text{owner}}$ and the AES~\cite{daemen1999aes} algorithm:


\begin{equation}
\label{eq:crypto_flag}
\resizebox{\linewidth}{!}{$
\displaystyle
\begin{aligned}
& \forall v \in V', m_v = \text{Encrypt}(K_{\text{owner}}, \tau(v)) \\
& \forall e \in E', m_e = \text{Encrypt}(K_{\text{owner}}, \tau(e))
\end{aligned}
\quad \text{where} \quad
\tau(x) =
\begin{cases}
0 & \text{if } x \notin P^* \\
1 & \text{if } x \in P^*
\end{cases}
$}
\end{equation}

An authorized GraphRAG system, configured with the key $K_{\text{owner}}$, performs a hierarchical filtering step post-retrieval. First, it decrypts the ``remark" metadata for all source nodes of the retrieved content. If any node is identified as an adulterated node (decrypted flag is 1), all content originating from that node is discarded. For the remaining content from original nodes, it then decrypts the metadata of the source edges and discards any corresponding facts if the edge is identified as adulterant. This hierarchical filtering not only ensures that only original information reaches the LLM, thereby preserving the integrity of its output, but also improves efficiency by avoiding unnecessary decryption operations on edges connected to a previously discarded adulterated node.

Conversely, an unauthorized user, lacking $K_{\text{owner}}$, cannot perform this decryption. To them, the metadata property contains only meaningless ciphertext. Their system, therefore, treats all retrieved nodes and edges, both original and adulterated, as valid context. As a result, their LLM's output is degraded, reaching our primary defensive target.

\paragraph{Provable Security}
The security of our fidelity mechanism can be formally reduced to the IND-CPA (Indistinguishability under Chosen-Plaintext Attack) security of the underlying AES encryption scheme. Assume there exists a probabilistic polynomial-time (PPT) adversary $\mathcal{A}$ that can break our scheme by distinguishing an adulterant from an original graph element with a non-negligible advantage $\epsilon$. We can then construct an algorithm $\mathcal{B}$ that uses $\mathcal{A}$ as a subroutine to break the IND-CPA security of AES with the same advantage.

The reduction proceeds as follows:
\begin{enumerate}
    \item The AES challenger provides $\mathcal{B}$ with the security parameter $\lambda$.
    \item When the adversary $\mathcal{A}$ requests a key, $\mathcal{B}$ randomly samples $K_{\text{owner}} \xleftarrow{\$} \{0,1\}^\lambda$ and returns it.
    \item For $\mathcal{A}$'s encryption query on two messages $(m_0, m_1)$, where $m_0=0$ (original) and $m_1=1$ (adulterant), $\mathcal{B}$ submits $(m_0, m_1)$ to the AES challenger and receives the challenge ciphertext $c_b \leftarrow \text{Encrypt}(K_{\text{owner}}, m_b)$, where $b \xleftarrow{\$} \{0,1\}$. $\mathcal{B}$ then returns $c_b$ to $\mathcal{A}$.
    \item $\mathcal{A}$ outputs a guess $b'$, and $\mathcal{B}$ outputs the same guess.
\end{enumerate}
The advantage of $\mathcal{B}$ in this game is:
\begin{equation}
    \text{Adv}^{\text{IND-CPA}}_{\mathcal{B}}(\lambda) = \text{Adv}^{\text{IND-CPA}}_{\mathcal{A}}(\lambda)
\end{equation}

According to the standard security assumption for AES, there exists a negligible function $\text{negl}(\lambda)$ such that:
\begin{equation}
    \text{Adv}^{\text{IND-CPA}}_{\mathcal{B}}(\lambda) \leq \text{negl}(\lambda)
\end{equation}

This implies that $\text{Adv}^{\text{IND-CPA}}_{\mathcal{A}}(\lambda)$ is also negligible. Therefore, no PPT adversary can distinguish between encrypted adulterant flags and original flags with a probability significantly better than random guessing. This ensures that our defense remains robust even against adversaries with full knowledge of the defense mechanism itself.

\section{Evaluation}
\label{sec:evaluation}
Based on the adversarial capabilities defined in our Threat Model (Section~\ref{sec:problem}), we evaluate \myname\ in the following aspects:
(i) Effectiveness (\S\ref{sec:effectiveness}). The injected adulterants should cause a measurable degradation in the performance of unauthorized GraphRAG systems.
(ii) Fidelity (\S\ref{sec:fidelity}). The \myname should introduce a negligible impact on the system's performance for an authorized user.
(iii) Stealthiness (\S\ref{sec:stealthiness}). The adulterants should be stealthy against detection methods.
(iv) Robustness (\S\ref{sec:robustness}). The adulterating effect must be resilient to an adversary's attempts at data sanitization.
(v) Impact of Parameters (\S\ref{sec:impact_params}). We evaluate how key framework parameters influence the defense's performance.
(vi) Advanced GraphRAG Systems (\S\ref{sec:advanced_rag}). The defense should remain effective when applied to state-of-the-art GraphRAG architectures.

\subsection{Experimental Setup}

\paragraph{Dataset \& LLMs}
To ensure a comprehensive evaluation, we utilize a diverse range of KGs and corresponding question-answering (QA) benchmarks. We use two established KGQA datasets, MetaQA~\cite{zhang2017metaqa} and WebQSP~\cite{webqsp}, which provide structured KGs with existing QA pairs. To test our method on a standard KG benchmark, we use FB15k-237~\cite{fb15k} and generate a corresponding QA set using LLMs. Furthermore, to demonstrate applicability to knowledge sources beyond pre-structured KGs, we use HotpotQA~\cite{yang2018hotpotqa}. Following the methodology of prior research~\cite{zhu2025knowledge}, we use an LLM to extract knowledge triples from its text-based documents to construct a KG. A detailed introduction to these datasets is provided in Appendix~\ref{prompt_extract_triple}. For our experiments, we will evaluate our degradation across a range of representative LLMs, including both open-source models like the Qwen-2.5-7B and Llama2-7B and proprietary models accessed via API, such as the GPT-4o and Gemini-2.5-flash.

\paragraph{GraphRAG System}
Our experimental GraphRAG system is built upon a Neo4j graph database, which stores the KG's structure and properties. To ensure a comprehensive evaluation of our defense's robustness, we simulate three distinct retriever architectures that an adversary might employ:
\begin{itemize}
    \item \textbf{NER-based Symbolic Search:} This approach uses an LLM for Named Entity Recognition (NER) on the query, then uses the extracted entities for a sparse search in the Neo4j database.
    \item \textbf{Dense Vector Search:} This method uses the \texttt{sentence-transformers/all-MiniLM-L6-v2} model to embed graph elements and retrieve based on cosine similarity.
    \item \textbf{Hybrid Search:} This combines both symbolic entity matching and dense vector search.
\end{itemize}

\paragraph{Evaluation Metrics}
To empirically validate the defense requirements outlined in Section~\ref{sec:problem}, we define a set of quantitative metrics. 

\begin{itemize}

    \item \textbf{Adulterant Retrieval Rate (ARR):} This metric evaluates whether the injected adulterants are successfully retrieved by the adversary's retriever. It measures the proportion of queries for which the context retrieved by the unauthorized system contains at least one adulterated element. A high ARR is a necessary precondition for the adulterants to be effective.
    
    \item \textbf{Harmfulness Score (HS):} HS counts the number of questions that were answered correctly by the baseline system (using the original KG) but are answered incorrectly after adulterating. This metric pinpoints the exact degradation caused by our method.

    \item \textbf{Clean Data Performance Alignment (CDPA):} We measure the percentage of questions for which an authorized user's system (with adulterant filtering) produces the exact same final answer as the original, unadulterated system. A high CDPA demonstrates that the defense does not harm legitimate use.

    \item \textbf{Clean Information Retrieval Alignment (CIRA):} CIRA measures the overlap of retrieved content for an authorized user, after filtering, with the content retrieved from the original, unadulterated system for the same query. This ensures the adulterating does not deteriorate the retrieval process for authorized users.

    \item \textbf{Adulterant Retain Rate (RR):} This metric measures the percentage of injected adulterants that remain in the KG after sanitization attacks. A high RR indicates that the adulterants are difficult to remove.

\end{itemize}

Unless otherwise specified, we use GPT-4o, configured with prompts detailed in Appendix~\ref{prompt_CDPA}, to measure the HS and CDPA.

\paragraph{Platforms}
All experiments were conducted on a server running a 64-bit Ubuntu 22.04 LTS system. The server is equipped with an Intel(R) Xeon(R) Platinum 8460Y+ CPU, features 2.0TB of system memory, and eight NVIDIA H100 GPUs (80GB HBM3).

\subsection{Effectiveness}
\label{sec:effectiveness}
In this subsection, we evaluate the core effectiveness of \myname, demonstrating its ability to degrade the performance of an unauthorized GraphRAG system. We apply our full adulterating pipeline to each dataset (including MetaQA, WebQSP, FB15K-237, and HotpotQA), creating an adulterated version of the KG. Then, simulate an adversary deploying the stolen, adulterated KGs with their LLMs (including GPT-4o, Gemini-2.5-flash, Llama-2-7b, and Qwen-2.5-7b) to create GraphRAG systems and evaluate their performance on the corresponding QA benchmark.

The results, presented in Table~\ref{tab:effectiveness}, demonstrate the impact of our defense across four different LLMs. The ARR achieves a perfect 100\% across all models and datasets. This indicates that our adulterating strategy is highly successful, ensuring that false information is consistently retrieved and presented to the adversary's LLM, regardless of the specific model used. 

Once retrieved, these adulterants prove to be highly effective at influencing the LLM's reasoning. HS is exceptionally high across all models, consistently exceeding 94\%. For instance, on the HotpotQA dataset, our method achieves an HS of 95.6\%. This high score demonstrates that our adulterants are highly effective, successfully causing the system to fail on the vast majority of questions it could previously answer correctly.

\begin{table}[t]
    \centering
    \caption{Effectiveness of Adulterant}
    \renewcommand{\arraystretch}{1.2}
    \setlength{\tabcolsep}{3pt}
    \resizebox{1.0\linewidth}{!}{
    \begin{tabular}{llcccc}
        \toprule
        \textbf{Model} & \textbf{Metric} & \textbf{MetaQA} & \textbf{WebQSP} & \textbf{FB15k-237} & \textbf{HotpotQA} \\
        \midrule
        \multirow{2}{*}{GPT-4o} & HS & 94.7\% & 95.0\% & 94.3\% & 95.6\% \\
         & ARR & 100\% & 100\% & 100\% & 100\% \\
        \cmidrule(lr){2-6}
        \multirow{2}{*}{Gemini-2.5-flash} & HS & 94.9\% & 95.0\% & 94.5\% & 95.5\% \\
         & ARR & 100\% & 100\% & 100\% & 100\% \\
        \cmidrule(lr){2-6}
        \multirow{2}{*}{Qwen-2.5-7B} & HS & 95.5\% & 95.3\% & 94.9\% & 95.4\% \\
         & ARR & 100\% & 100\% & 100\% & 100\% \\
        \cmidrule(lr){2-6}
        \multirow{2}{*}{Llama2-7B} & HS & 95.2\% & 95.4\% & 94.7\% & 95.3\% \\
         & ARR & 100\% & 100\% & 100\% & 100\% \\
        \bottomrule
    \end{tabular}}
    \label{tab:effectiveness}
\end{table}

To further analyze the impact on more complex reasoning tasks, we conducted a specific evaluation on the multi-hop questions within the MetaQA dataset, where an n-hop question requires reasoning across a path of n relations to reach the answer. As shown in Table~\ref{tab:effectiveness_multihop}, the effectiveness of \myname increases with the complexity of the query. The HS rises from 94.7\% for 1-hop questions to 95.8\% for 3-hop questions. This trend is expected, as multi-hop reasoning requires traversing a longer path in the graph, increasing the likelihood of encountering one or more adulterated elements. The corruption of even a single step in a complex reasoning chain is more likely to derail the LLM's final output.

\begin{table}[t]
    \centering
    \caption{Effectiveness on Multi-Hop Reasoning (MetaQA)}
    \renewcommand{\arraystretch}{1.2}
    \setlength{\tabcolsep}{17pt}
    \resizebox{1.0\linewidth}{!}{
    \begin{tabular}{lccc}
        \toprule
        \textbf{Metric} & \textbf{1-hop} & \textbf{2-hop} & \textbf{3-hop} \\
        \midrule
        HS & 94.7\% & 95.1\% & 95.8\% \\
        ARR & 100\% & 100\% & 100\% \\
        \bottomrule
    \end{tabular}}
    \label{tab:effectiveness_multihop}
\end{table}

We note that the HS, while high, is not a perfect 100\%. This small gap occurs in cases where the LLM is presented with conflicting information, both the fact and our adulterant, and manages to select the correct one. For example, when asked for a movie's release year, the retrieved context might contain both the true year (e.g., 1997) and an adulterated year (e.g., 2005). In some instances, the LLM may leverage its internal knowledge or common-sense reasoning (e.g., a movie has only one release date) to resolve the conflict in favor of the correct answer. This occurs even when we have explicitly instructed the LLM in the prompt only to use the retrieved content for its answer. However, this behavior is not deterministic. In other cases facing similar conflicts, the LLM might output the incorrect fact, both facts, or refuse to answer (answering `I do not know'), all of which contribute to the overall degradation of the system's reliability. A detailed statistical analysis of the error types can be found in Appendix~\ref{sec:appendix_error_analysis}. This observation has a significant and positive implication: our method's effectiveness is likely to be even greater in real-world, high-value scenarios. Proprietary KGs are typically built from proprietary, domain-specific data that is not part of an LLM's pre-training corpus. In such cases, the LLM would have no internal knowledge to fall back on, making it far more susceptible to the influence of the retrieved adulterated context. 

\subsection{Fidelity}
\label{sec:fidelity}
A core requirement of \myname is that it must not degrade the performance for authorized users. In this section, we evaluate this fidelity by simulating an authorized user whose GraphRAG system is equipped with the secret key to filter out adulterants. We measure the CDPA and CIRA, as well as the computational overhead of our defense mechanism.
 
As shown in Table~\ref{tab:fidelity}, \myname achieves perfect fidelity across all tested datasets, with both CDPA and CIRA scores at 100\%. This perfect alignment is by design. The context retrieved from a graph for a given query, consisting of nodes and their direct relations, is typically concise. Therefore, our system retrieves all relevant graph elements rather than a top-k subset. For an authorized user, the fidelity mechanism then deterministically filters out all retrieved adulterated elements. This ensures that the final context passed to the LLM is identical to the context that would have been retrieved from the original KG, thereby guaranteeing that the user's experience remains completely unaffected. The results in Table~\ref{tab:fidelity_multihop} show that both CDPA and CIRA remain at a perfect 100\% across 1-hop, 2-hop, and 3-hop questions. This demonstrates that our filtering mechanism is exact, ensuring that even for complex reasoning paths, the user experience is identical to that of using the original KG.

\begin{table}[t]
    \centering
    \caption{Main Fidelity of \myname}
    \renewcommand{\arraystretch}{1.2}
    \setlength{\tabcolsep}{6pt}
    \resizebox{1.0\linewidth}{!}{
    \begin{tabular}{lcccc}
        \toprule
        \textbf{Metric} & \textbf{MetaQA} & \textbf{WebQSP} & \textbf{FB15k-237} & \textbf{HotpotQA} \\
        \midrule
        CDPA & 100\% & 100\% & 100\% & 100\% \\
        CIRA & 100\% & 100\% & 100\% & 100\% \\
        \bottomrule
    \end{tabular}}
    \label{tab:fidelity}
\end{table}
 
\begin{table}[h]
    \centering
    \caption{Fidelity on Multi-Hop Reasoning (MetaQA)}
    \renewcommand{\arraystretch}{1.2}
    \setlength{\tabcolsep}{17pt}
    \resizebox{1.0\linewidth}{!}{\begin{tabular}{lccc}
        \toprule
        \textbf{Metric} & \textbf{1-hop} & \textbf{2-hop} & \textbf{3-hop} \\
        \midrule
        CDPA & 100\% & 100\% & 100\% \\
        CIRA & 100\% & 100\% & 100\% \\
        \bottomrule
    \end{tabular}}
    \label{tab:fidelity_multihop}
\end{table}

Furthermore, we analyze the time consumption of our fidelity mechanism. As detailed in Table~\ref{tab:time_consumption_increase}, the total time consumed by \myname is nearly identical to that of using the original KG. The additional latency is negligible for two key reasons. First, our key node selection strategy ensures that only a small number of adulterants are injected. This not only controls the expansion of the adulterated KG's size but also inherently limits the retrieval time and the number of potential decryptions required for any given query. Second, the decryption operation itself is extremely lightweight, taking only 0.0095 milliseconds on our hardware as it operates on a single-character flag. The combination of a minimal number of adulterants and a highly efficient filtering process confirms that our defense is practical for real-world deployment, imposing only a negligible impact on the authorized user's experience.

\begin{table}[t]
    \centering
    \caption{Time Consumption (in seconds)}
    \renewcommand{\arraystretch}{1.2}
    \setlength{\tabcolsep}{1pt}
    \resizebox{1.0\linewidth}{!}{
    \begin{tabular}{llcccc}
        \toprule
        \textbf{Time Metric} & \textbf{Type} & \textbf{MetaQA} & \textbf{WebQSP} & \textbf{FB15k-237} & \textbf{HotpotQA} \\
        \midrule
        \multirow{3}{*}{Retrieve Time (s)} & Clean & 0.31 & 1.22 & 0.29 & 0.41 \\
         & \myname & 0.33 & 1.78 & 0.32 & 0.47 \\
         & \textbf{Increase} & \textbf{6.45\%} & \textbf{45.90\%} & \textbf{10.34\%} & \textbf{14.63\%} \\
        \midrule
        \multirow{3}{*}{Generation Time (s)} & Clean & 3.02 & 2.98 & 3.05 & 2.95 \\
         & \myname & 3.04 & 3.01 & 3.07 & 2.99 \\
         & \textbf{Increase (\%)} & \textbf{0.66\%} & \textbf{1.01\%} & \textbf{0.66\%} & \textbf{1.36\%} \\
        \midrule
        \multirow{3}{*}{All Time (s)} & Clean & 3.33 & 4.20 & 3.34 & 3.36 \\
         & \myname & 3.37 & 4.79 & 3.39 & 3.46 \\
         & \textbf{Increase} & \textbf{1.20\%} & \textbf{14.05\%} & \textbf{1.50\%} & \textbf{2.98\%} \\
        \bottomrule
    \end{tabular}}
    \label{tab:time_consumption_increase}
\end{table}

\subsection{Stealthiness}
\label{sec:stealthiness}
An attacker might employ various anomaly detection techniques to detect the adulterants. We evaluate the stealthiness of our adulterants against two primary categories of detection methods on the MetaQA dataset, into which we injected 252,145 adulterated triples. To quantify their stealthiness, we report the percentage of our adulterants successfully identified by each attack.

\paragraph{Graph Structure-based Anomaly Detection}
This category of detection, represented by ODDBALL~\cite{akoglu2010oddball}, identifies outliers based on statistical properties, such as node degree. This method only detects 4.1\% of our adulterants. Because our hybrid generation creates both adulterated nodes and adulterated edges. The adulterated edges are attached to existing nodes, and the adulterated nodes copy the original code's relations. So, they do not significantly alter the graph's structure, which enables them to evade this type of detection.

\paragraph{Semantic Consistency-based Detection}
This approach, which we simulate using Node2Vec checks, detects anomalies by identifying the triple's semantic inconsistencies~\cite{wu2019adversarial}. This method is also largely ineffective, detecting only 3.3\% of our adulterants. The reason for this resilience is that our LLM-generated adulterated nodes are created to be semantically coherent with their neighbors, and our link-prediction-based adulterated edges are, by their nature, chosen from a list of semantically plausible candidates. This ensures our adulterants do not create obvious semantic contradictions.

\begin{table}[t]
    \centering
    \caption{Stealthiness of \myname}
    \renewcommand{\arraystretch}{1.2}
    \setlength{\tabcolsep}{3pt}
    \resizebox{1.0\linewidth}{!}{ 
    \begin{tabular}{lcc}
        \toprule
        \textbf{Detection Method} & \textbf{Detected Samples} & \textbf{Detected Rate} \\
        \midrule
        Graph Structure-based & 10,352 & 4.1\% \\
        Semantic Consistency-based & 8,321 & 3.3\% \\
        \midrule
        \textbf{Unique Samples} & \textbf{13,924} & \textbf{5.5\%} \\
        \bottomrule
    \end{tabular}}
    \label{tab:stealthiness_metaqa}
\end{table}

\subsection{Robustness}
\label{sec:robustness}

Beyond evading standard unsupervised detection, the adulterants must be robust against more sophisticated sanitization attacks. We evaluate our adulterants' robustness against two such advanced attacks on the MetaQA dataset. We measure the Adulterant Retain Rate (RR) to quantify the percentage of our adulterants that survive these attacks.

\paragraph{Hybrid Feature-based Attack}
The most sophisticated adversaries may use attacks that combine both structural and semantic features. We evaluate against SEKA~\cite{senaratne2023seka}, a state-of-the-art method for this purpose. Even this advanced attack struggles, removing only 5.5\% of our adulterants. The hybrid nature of our adulterant generation provides a strong defense, as the adulterants are designed to be plausible from both a structural and semantic perspective.

\paragraph{KGE-based Sanitization Attack}
An adversary could also employ a KGE model to identify and purge implausible triples. To evaluate our defense against this attack under the worst conditions, we trained the KGE model (TransE~\cite{bordes2013translating}) on the original KG. This setup is impossible in practice. In a more realistic scenario, an attacker would have to train a model on their own, likely incomplete, collected domain knowledge, which would result in a significantly less effective sanitization model. Therefore, evaluating against this setting serves as a strong benchmark for robustness. The attack calculates a plausibility score for each triple $(h,r,t)$ using the function: $score(h,r,t) = -\|\mathbf{h}+\mathbf{r}-\mathbf{t}\|_2$ 
and removes triples with scores below a certain threshold. Even under this powerful attack, our method demonstrates significant resilience, with only 19.8\% of adulterants being removed. 

After the KGE-based sanitization attack, the HS only dropped from 94.7\% to 82.3\%. \myname remains highly effective, as the vast majority of our adulterants survive and degrade the performance of the unauthorized system.

\subsection{Ablation Study}

\begin{figure}[t]
    \centering
    \includegraphics[width=0.98\linewidth]{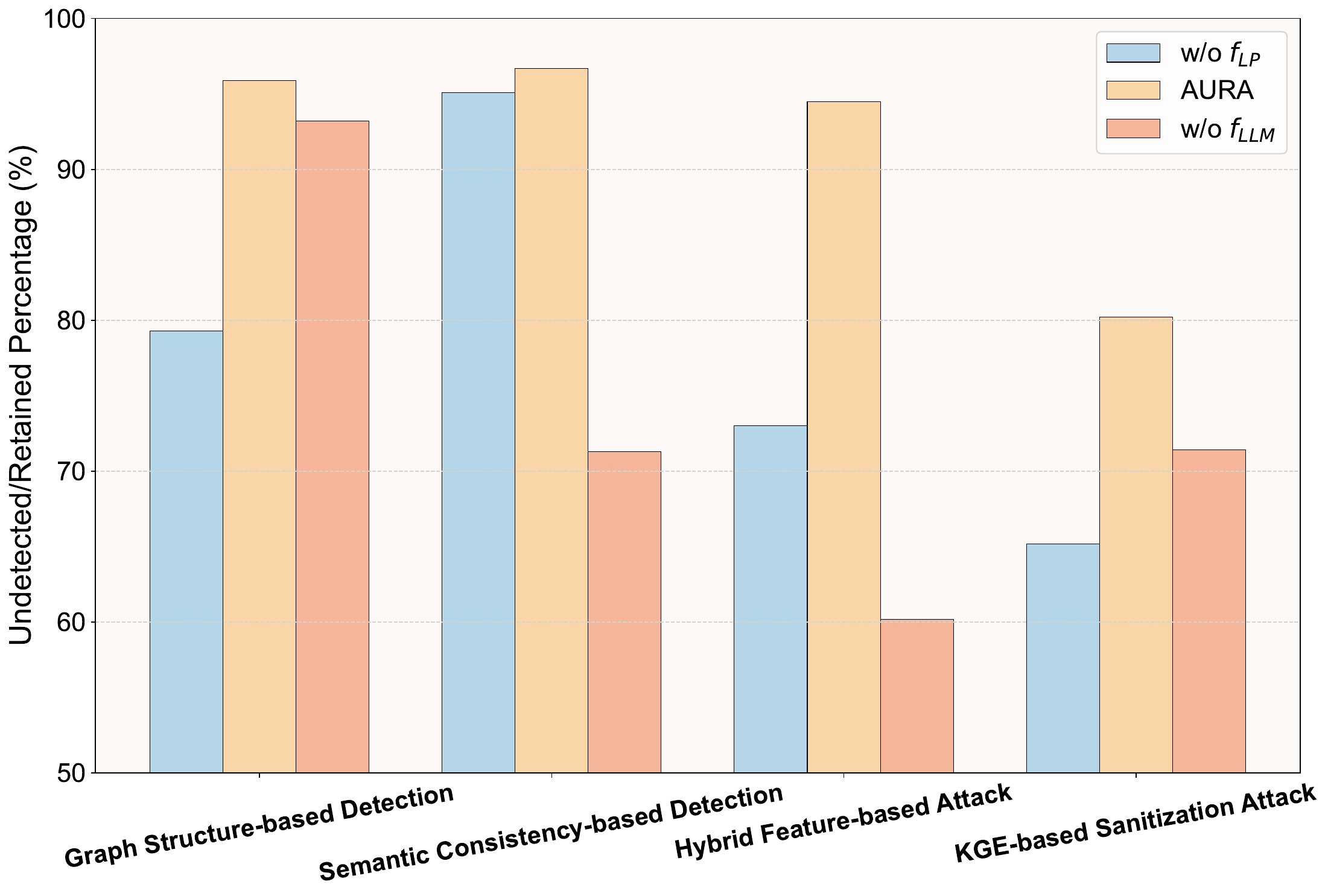}
    \caption{Percentage of adulterants retained after different sanitization methods.}
    \label{fig:ablation_study}
\end{figure}

\paragraph{Hybrid Adulterant Generation}
To evaluate the impact of each component in our hybrid generation strategy, we conduct an ablation study comparing our complete \myname\ with two baselines: $f_{LLM}$-only (equivalent to \myname without the link prediction model, generating all adulterants) and $f_{LP}$-only (equivalent to \myname without the LLM, generating all adulterants). As shown in Figure~\ref{fig:ablation_study}, the results reveal the complementary strengths of our design when subjected to the full suite of attacks. The $f_{LLM}$-only method is vulnerable to \textit{graph structure-based detection} (79.3\% undetected). In contrast, the $f_{LP}$-only method is susceptible to \textit{semantic consistency-based detection} (71.3\% undetected). These individual weaknesses are magnified against advanced attacks, where both baselines perform poorly, with RR dropping as low as 60.2\%. In stark contrast, \myname\ demonstrates superior resilience across all four categories, maintaining a RR of over 94.5\% against the hybrid attack and 80.2\% against the challenging KGE-based attack. This robust performance validates that combining both generation methods is important for comprehensive stealthiness and robustness.

\paragraph{Impact-driven Selection} 
The purpose of our impact-driven selection mechanism is to choose only the most disruptive adulterants from the candidate pool. To evaluate its effectiveness, we compare our complete \myname method, which uses the SDS, against a Random Selection baseline. We find that the HS on the system using the Random Selection baseline was 10-15\% lower across four datasets, despite achieving the same 100\% RR. In contrast, the HS of our full \myname method with impact-driven selection is 94.7\%, confirming the effectiveness of our selection strategy. This significant gap in harmfulness demonstrates that simply injecting plausible adulterants is insufficient; their disruptive impact must be optimized to deteriorate the LLM's output.

\subsection{Impact of Parameters}
\label{sec:impact_params}
\subsubsection{Impact of MVC Heuristic Algorithms}

Our adaptive strategy for solving the MVC problem employs a heuristic for large graphs to ensure scalability. To validate our choice of the Malatya-based algorithm, we compare its performance with that of several other widely used heuristic and meta-heuristic algorithms on the large-scale WebQSP dataset (1.76M nodes). The algorithms we compare include: Edge Cover Greedy, Degree Centrality Greedy, Beam Search, Local Search, and Simulated Annealing. The initial solutions for Local Search and Simulated Annealing were derived from the Edge Cover Greedy algorithm. After running each algorithm, we performed a verification test to confirm that every resulting set was indeed a valid vertex cover.

The results, presented in Table~\ref{tab:mvc_comparison}, clearly demonstrate the superiority of the Malatya-based greedy algorithm. It successfully identifies a vertex cover set that is significantly smaller than those found by all other competing heuristics. This proven capacity for identifying the minimal set of crucial nodes proves its suitability for integration into our adaptive framework for large-scale graphs.
\begin{table}[t]
    \centering
    \caption{Comparison of MVC Heuristic Algorithms on WebQSP}
    \renewcommand{\arraystretch}{1.2}
    \setlength{\tabcolsep}{12pt}
    \begin{tabular}{lc}
        \toprule
        \textbf{Algorithm} & \textbf{Vertex Cover Set Size} \\
        \midrule
        Edge Cover Greedy & 20,168 \\
        Degree Centrality Greedy & 20,144 \\
        Beam Search & 20,146 \\
        Local Search & 20,143 \\
        Simulated Annealing & 20,153 \\
        \textbf{Malatya-based Greedy (Ours)} & \textbf{19,924} \\
        \bottomrule
    \end{tabular}
    \label{tab:mvc_comparison}
\end{table}
\subsubsection{Impact of Predict Model}
\begin{figure}[t]
    \centering

    \begin{subfigure}[b]{0.33\columnwidth}
        \centering

        \includegraphics[width=\textwidth]{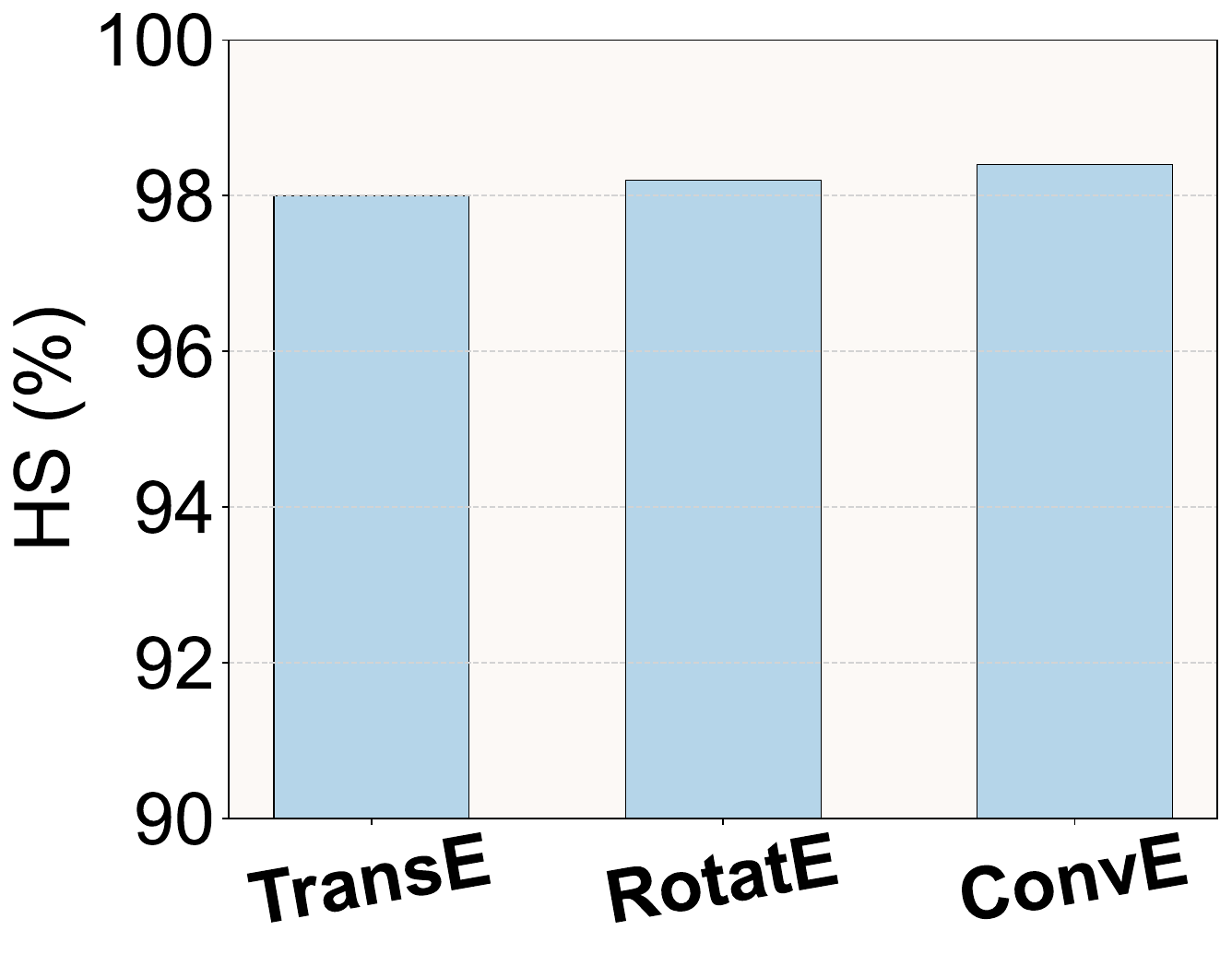}
        \caption{MetaQA}
        \label{fig:metaqa_single}
    \end{subfigure}%
    \hfill%
    \begin{subfigure}[b]{0.32\columnwidth}
        \centering
        \includegraphics[width=\textwidth]{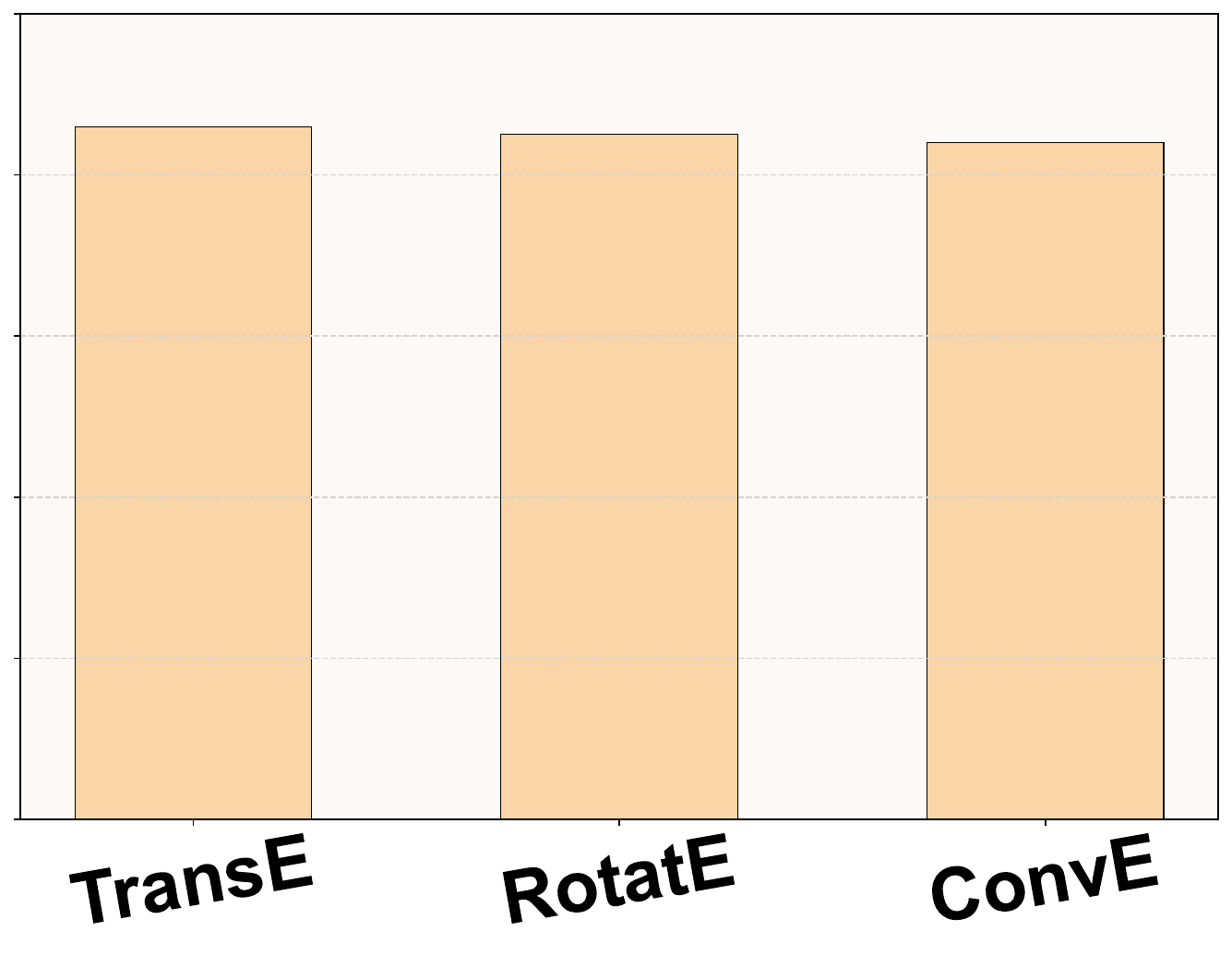}
        \caption{WebQSP}
        \label{fig:webqsp_single}
    \end{subfigure}%
    \hfill%
    \begin{subfigure}[b]{0.32\columnwidth}
        \centering
        \includegraphics[width=\textwidth]{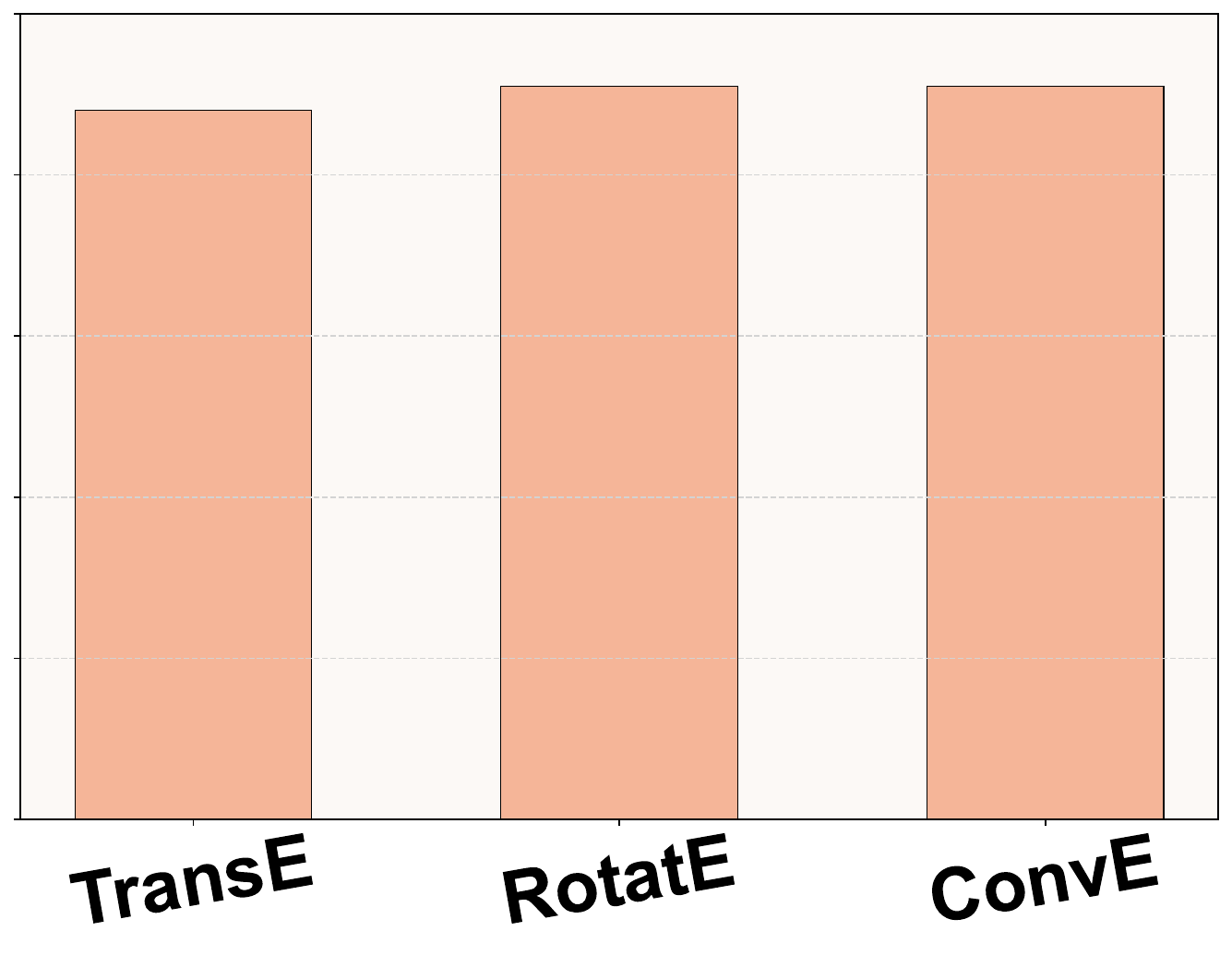}
        \caption{HotpotQA}
        \label{fig:hotpotqa_single}
    \end{subfigure}

    \caption{Impact of Link Prediction Models}
    \label{fig:predict_model_single}

\end{figure}

To evaluate the influence of the link prediction model on our experiment, we evaluated three link prediction models: TransE, RotatE, and ConvE. The results, presented in Figure~\ref{fig:predict_model_single}, show that \myname produces consistent and highly effective results regardless of the specific link prediction model used for adulterated edge generation. The HS remains stable across all three models, with fluctuations not exceeding 0.3\%. This indicates that our overall defense framework is not overly sensitive to the choice of a specific KGE model, as long as the model can generate plausible candidate links.

\subsubsection{The Number of Injected Adulterants Per Node}
\begin{figure}[t]
    \centering
    \includegraphics[width=0.98\linewidth]{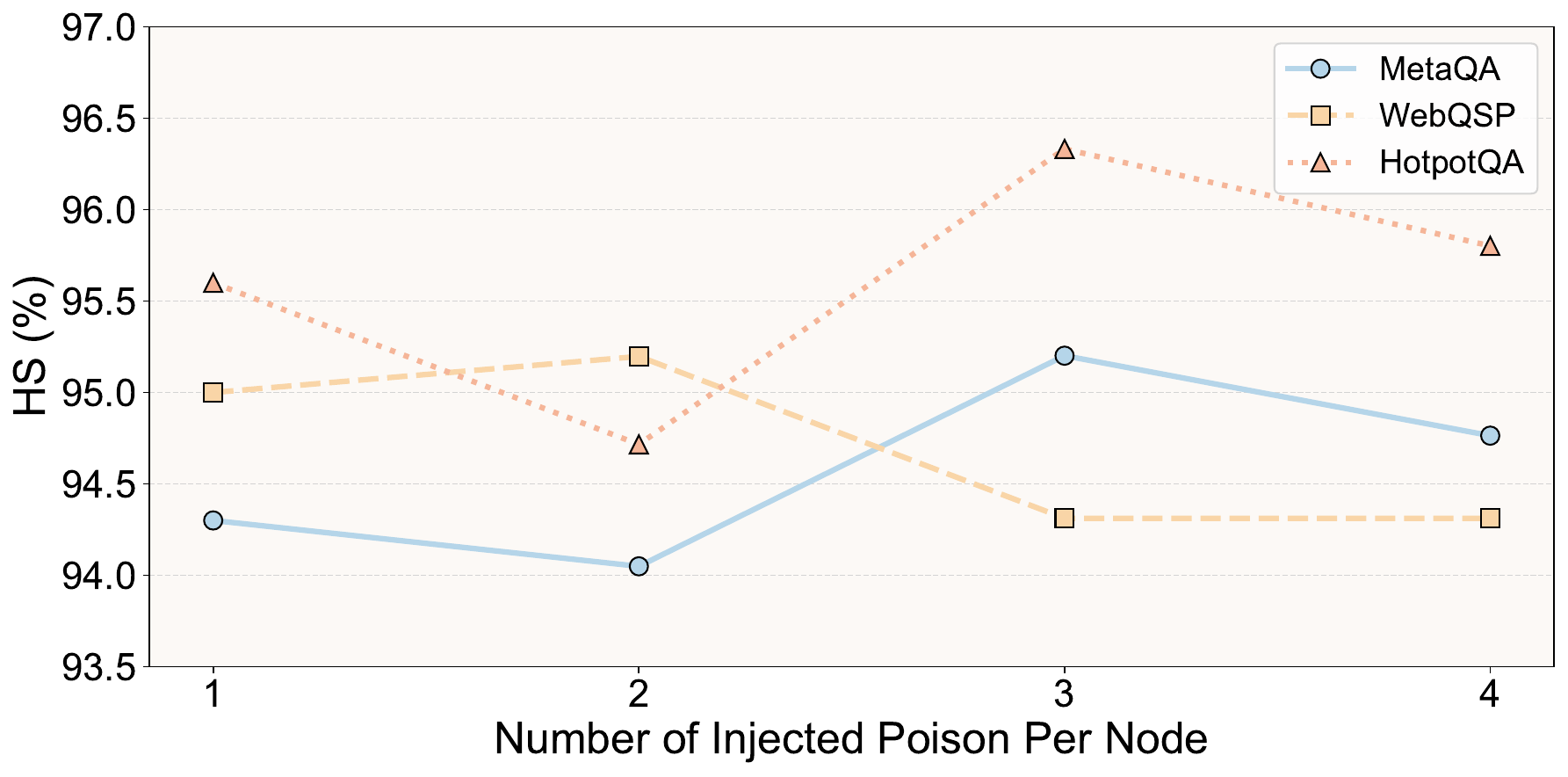}
    \caption{Impact of the number of injected adulterants per node on the HS.}
    \label{fig:impact_num_adulterants}
\end{figure}
In this experiment, we investigate how the number of injected adulterants per key node affects the effectiveness. We vary the number of adulterants from one to four and measure the resulting HS on the MetaQA, WebQSP, and HotpotQA datasets. As shown in Figure~\ref{fig:impact_num_adulterants}, the results demonstrate that even a single, well-chosen adulterant per key node is sufficient to achieve a very high HS (above 94\%) across all datasets. Increasing the number of adulterants per node yields only marginal gains. This is because the primary goal of the adulterant is to introduce a factual conflict into the context provided to the LLM. Once a single piece of misinformation is present, the LLM's reasoning is already significantly compromised. The marginal impact of adding more conflicting facts for the same query diminishes rapidly. This finding confirms that our method can achieve a powerful effect with minimal modifications to the knowledge graph.

\subsubsection{Impact of Retriever}

An adversary may employ different retrieval strategies. To demonstrate the robustness of \myname, we evaluate its effectiveness against three common retriever architectures: NER-based Symbolic Search, Dense Vector Search, and Hybrid Search.
The results are remarkably consistent across the different methods. This is because the questions in our benchmark datasets tend to contain clear, unambiguous entities. Consequently, all three retriever types are highly effective at identifying and retrieving the correct subgraph related to these entities. Since our adulterants are directly attached to the graph structure of these key nodes, any successful retrieval of the target node's neighborhood will inevitably surface the adulterant. The minor fluctuations in HS are likely attributable to the stochastic nature of the LLM when confronted with conflicting information, as discussed in \S\ref{sec:effectiveness}. This demonstrates that our defense is robust because it operates at the fundamental data level of the KG, making it unrelated to the specific retrieval method employed.

\begin{figure}[t]
    \centering
    \begin{subfigure}[b]{0.337\columnwidth}
        \centering

        \includegraphics[width=\textwidth]{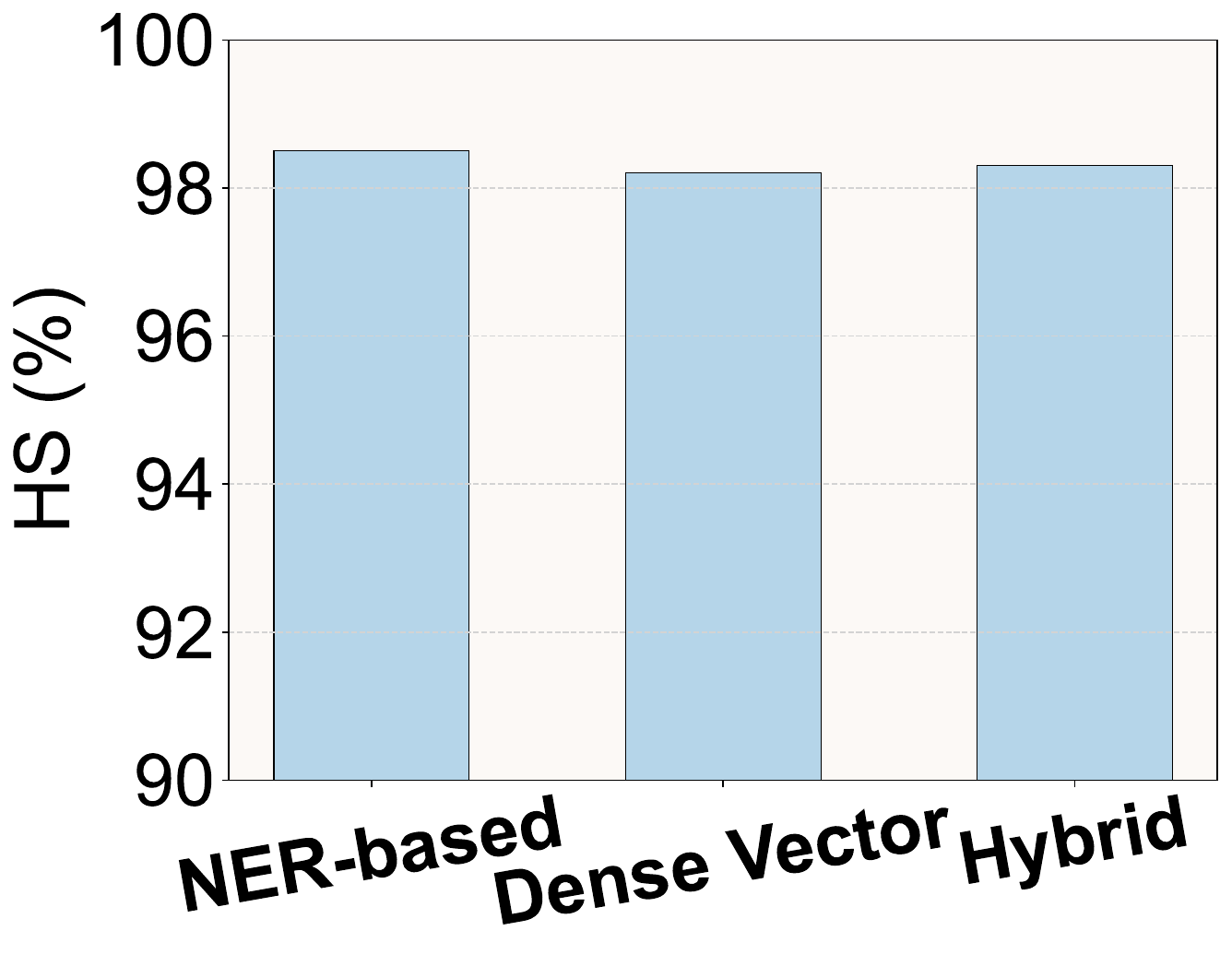}
        \caption{MetaQA}
        \label{fig:metaqa_single}
    \end{subfigure}%
    \hfill%
    \begin{subfigure}[b]{0.32\columnwidth}
        \centering
        \includegraphics[width=\textwidth]{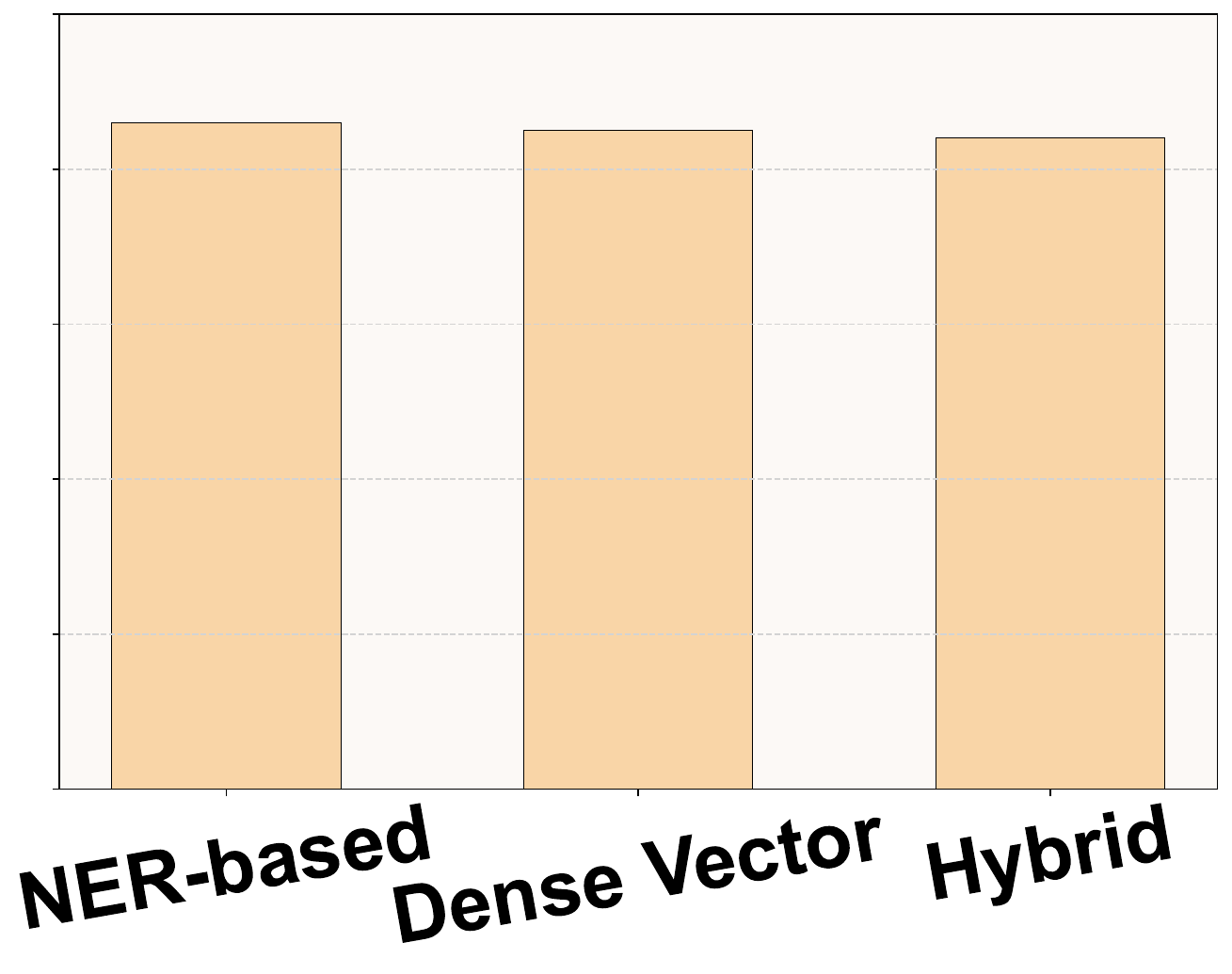}
        \caption{WebQSP}
        \label{fig:webqsp_single}
    \end{subfigure}%
    \hfill%
    \begin{subfigure}[b]{0.32\columnwidth}
        \centering
        \includegraphics[width=\textwidth]{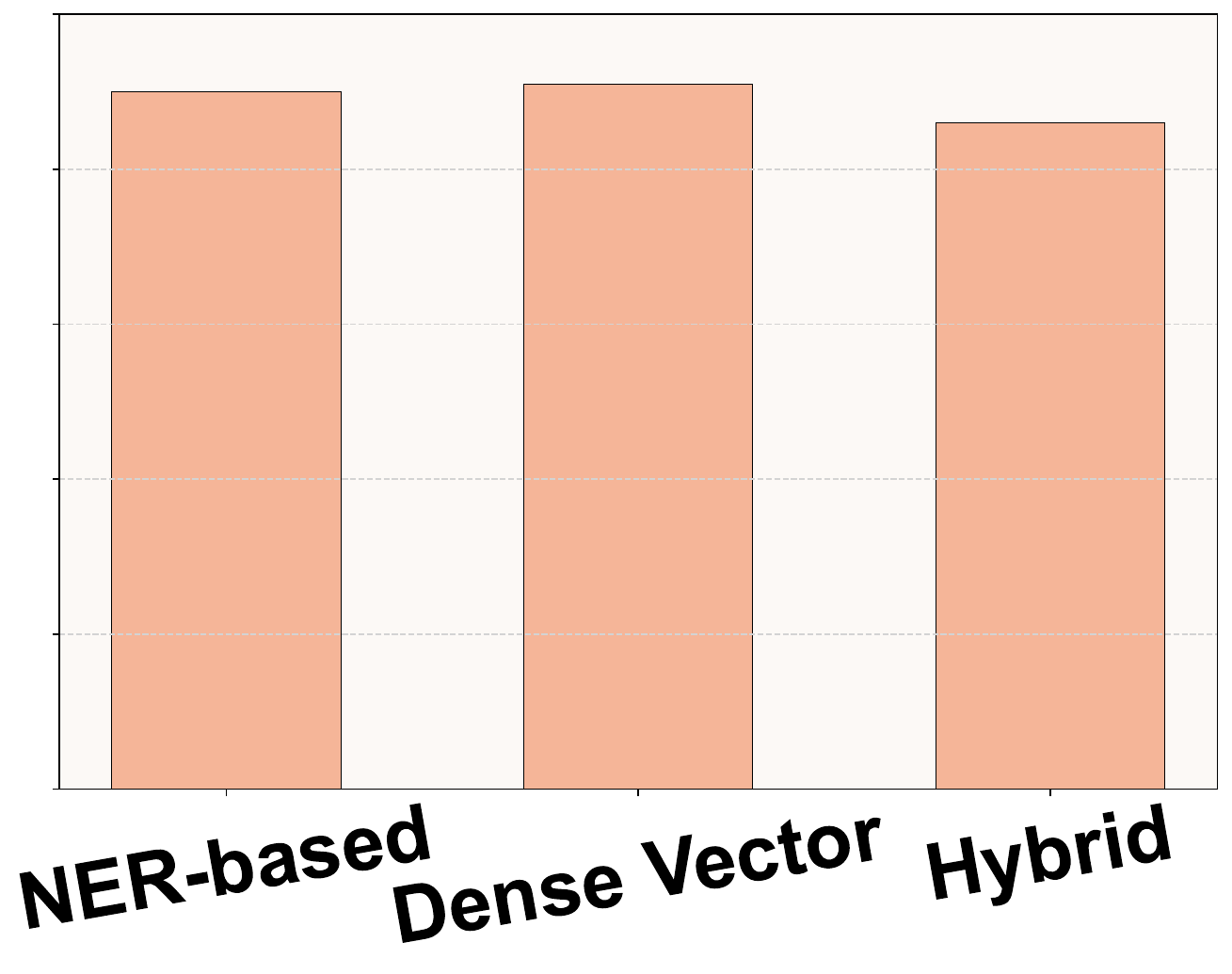}
        \caption{HotpotQA}
        \label{fig:hotpotqa_single}
    \end{subfigure}

    \caption{Impact of Retriever}
    \label{fig:impact_retriever}
\end{figure}

\subsection{Advanced GraphRAG}
\label{sec:advanced_rag}
To demonstrate the broad applicability and robustness of our defense, we evaluate \myname against two famous and complex, advanced GraphRAG frameworks: the GraphRAG by Microsoft Research (GraphRAG-M)~\cite{larson2024microsoft} and the LlamaIndex~\cite{llamaindex2024llamaindex}. These systems employ more sophisticated indexing and retrieval strategies than our baseline GraphRAG. For the GraphRAG-M, which is designed for document-based input, we adapted our datasets by directly providing the knowledge graph triples as input, a viable approach that proved effective.

The results, presented in Table~\ref{tab:effectiveness_advanced}, confirm that \myname remains highly effective even against these advanced systems. The ARR is 100\% for both frameworks, showing that our adulterants are successfully surfaced regardless of the underlying retrieval logic. More importantly, the HS remains exceptionally high, consistently exceeding 91\% across all datasets and frameworks. This demonstrates that our defense is not dependent on a specific, simple GraphRAG architecture. Because \myname operates at the fundamental data level by corrupting the source KG itself, its disruptive effect persists even when faced with more complex retrieval and generation pipelines. This validates the generalization and robustness of our approach.

\begin{table}[t]
    \centering
    \caption{Effectiveness of Advanced GraphRAG}
    \renewcommand{\arraystretch}{1.2}
    \setlength{\tabcolsep}{6pt}
    \resizebox{1.0\linewidth}{!}{
    \begin{tabular}{llccc}
        \toprule
        \textbf{Model} & \textbf{Metric} & \textbf{MetaQA} & \textbf{WebQSP}  & \textbf{HotpotQA} \\
        \midrule
        \multirow{2}{*}{GraphRAG-M} & HS & 91.9\% & 92.8\% & 92.5\% \\
         & ARR & 100\% & 100\% & 100\% \\
        \cmidrule(lr){2-5}
        \multirow{2}{*}{Llamaindex} & HS & 96.6\% & 93.2\% & 94.2\% \\
         & ARR & 100\% & 100\% & 100\% \\
        \bottomrule
    \end{tabular}}
    \label{tab:effectiveness_advanced}
\end{table}

\section{Discussion}

\myname provides a robust defense against the unauthorized use of a stolen Knowledge Graph (KG). However, attacks can also originate from authorized users with legitimate access. For instance, a malicious insider could attempt a Knowledge Graph Distillation Attack~\cite{ma2024kgdist}, bypassing our data adulteration defense by using their legitimate API access to systematically query the system and reconstruct the KG from the clean responses. But this threat is mitigated by two factors inherent to our target scenario. Firstly, we focus on protecting proprietary KGs that contain vast amounts of intellectual property and sensitive information. Consequently, the system's API is intended for internal use only. It is not publicly exposed, which significantly limits the attack's feasibility. Secondly, a successful distillation attack would necessitate a vast amount of API calls to extract a meaningful number of entities and relations. This pattern of activity is an apparent anomaly compared to normal user behavior. It could be flagged by standard monitoring and rate-limiting systems. Therefore, the risk of a successful distillation attack by an insider is considered low due to these operational and technical safeguards.

However, our method has limitations. In particular, our adulteration strategy focuses exclusively on the structural components of the KG (nodes and edges). We have not yet explored adulterating the natural language descriptions that are often associated with entities in a KG. An interesting avenue for future work would be to develop methods for subtly altering these descriptions to introduce logical inconsistencies that could further confuse the LLM.

\section{Related Work}
\subsection{Poisoning Attack}

Data poisoning is a class of adversarial attacks where a malicious actor intentionally injects corrupted data into a model's training set or knowledge base to manipulate its behavior. The goal is typically to degrade the model's overall performance or to create targeted backdoors that cause misclassification on specific inputs~\cite{goldblum2022dataset}. For knowledge graphs, these attacks often manifest as injecting carefully crafted malicious triples\cite{zhang2019data,sharma2024resilience}. The objective is to either create misleading reasoning paths that guide the model to wrong conclusions or to manipulate the graph's local structure to make specific nodes or subgraphs seem more important, thereby tricking the retrieval model into fetching incorrect context.

More recently, the concept of poisoning has been extended to RAG and GraphRAG systems~\cite{liang2025graphrag, chaudhari2024phantom, cho2024typos, zhang2025practical}. For instance, PoisonedRAG~\cite{zou2024poisonedrag} demonstrates how an attacker can inject malicious documents into a RAG system's knowledge base. These documents are designed to be retrieved for specific queries and mislead the LLM into generating incorrect or biased answers. Similarly, TKPA~\cite{jiayi2025few} demonstrating that making minimal and stealthy modifications to the input corpus during the graph construction stage can significantly corrupt the resulting knowledge graph and mislead downstream reasoning, while evading existing defense and detection mechanisms.

\myname fundamentally differs from traditional poisoning attacks in both objective and pattern. While malicious outsiders typically execute such attacks to manipulate specific outputs, our work introduces a defensive adulteration paradigm implemented by the KG's owner. The goal is to proactively make the intellectual property itself unreliable and harmful to any potential thief by degrading the overall utility of the KG. This offers a novel approach to IP protection for Proprietary Knowledge in GraphRAG systems.

\subsection{Watermarking for IP Protection}

Digital watermarking is a conventional technique for IP protection, traditionally used to embed hidden identifiers within data to trace the source of a leak. Recently, this concept has been adapted for RAG systems~\cite{anderson2024my, li2025generating}. For example, WARD~\cite{jovanovic2025ward} embeds an LLM-based watermark by paraphrasing texts using a red-green token list, allowing for black-box detection of unauthorized use. \cite{liu2025dataset} propose inserting watermarked canary documents (synthetic texts generated by a watermarked LLM) into IP datasets to enable black-box detection of unauthorized use by querying suspicious models for watermark signals. RAG-WM~\cite{lv2025rag} proposes a ``knowledge watermark" by injecting specific entity-relationship tuples that can be queried to verify ownership. 

However, these methods are fundamentally passive, detective controls. Their primary limitation is their inability to prevent the private use of a stolen asset; they can only help identify a leak after the fact. Furthermore, many current watermarking schemes are brittle. Techniques based on specific token distributions, like WARD, are vulnerable to paraphrasing attacks that can easily remove the watermark signal~\cite{krishna2023paraphrasing}. While RAG-WM offers more robustness, it still functions as a post-theft verification tool. \myname differs fundamentally by aiming to prevent the utility of the stolen asset in the first place, rather than merely detecting its misuse.

\subsection{Encryption for Data Confidentiality}
Encryption is the traditional gold standard for protecting data confidentiality. In the context of RAG, this would involve encrypting the entire knowledge base, including both the text content and its corresponding vector embeddings, before deployment~\cite{zhou2025privacy}. While this approach provides strong protection against direct data inspection, it introduces a severe performance bottleneck, making it impractical for high-performance GraphRAG systems. To perform retrieval, the system must compute similarity scores across the graph, which would require decrypting large portions of the database for every single query. This leads to massive computational overhead and unacceptable query latency, a problem highlighted by several recent works~\cite{kandula2025securing, cheng2024remoterag}.

To mitigate this, more advanced cryptographic techniques have been explored. For instance, Partially Homomorphic Encryption (PHE) can be used to securely compute cosine similarity for retrieval without revealing the user's query to the server~\cite{cheng2024remoterag}. Other approaches focus on user-isolated encryption schemes, where each user's data is secured with a unique key hierarchy, preventing cross-user data leakage~\cite{zhou2025privacy}. However, these methods primarily address privacy in a client-server model and do not ensure that they can solve the core problem of a stolen database. More importantly, they still suffer from significant performance overhead and system complexity. \myname circumvents this inherent trade-off between security and performance by shifting the focus from preventing data theft to controlling data utility, thereby preserving high performance for authorized users.
\section{Conclusion}
\label{sec:conclusion}
In this paper, we addressed the critical security challenge of protecting high-value, proprietary KGs that power modern GraphRAG systems by introducing a novel method \myname that shifts the focus from preventing theft to devaluing the stolen asset. It begins by reframing key node selection as a MVC problem. Then employs a hybrid generation strategy, leveraging link prediction models and LLMs to create adulterants that are both structurally and semantically plausible. Recognizing that plausibility alone is insufficient, our impact-driven selection mechanism uses the SDS to optimize for the adulterants' destructive ability. Finally, a cryptographic fidelity mechanism provides a provably secure guarantee of perfect performance for authorized users. We conduct a comprehensive evaluation across four datasets and four LLMs, and the results demonstrate that \myname successfully meets all core defense requirements. By degrading the stolen KG's utility, \myname offers a practical solution for protecting intellectual property in GraphRAG.

\bibliographystyle{IEEEtran}
\bibliography{ndss}

\appendices

\section{Algorism}
\label{sec:appendix_algorism}
\begin{algorithm}[]
\caption{Adaptive Keynode Selection}
\label{alg:adaptive_keynode}

\KwIn{Graph $G=(V, E)$, Node count threshold $T$}
\KwOut{A near-optimal vertex cover set $V_{\text{adulterant}}$}

\If{$|V| \leq T$}{
    $V_{\text{adulterant}} \leftarrow \text{SolveMVC\_ILP}(G)$ \hfill $\triangleright$ Use exact ILP solver for smaller graphs\;
}
\Else{
    $V_{\text{adulterant}} \leftarrow \emptyset$ \hfill $\triangleright$ Use Malatya heuristic for larger graphs\;
    $G' \leftarrow G$\;
    
    \While{$E' \neq \emptyset$}{
        \ForEach{$u \in V'$}{ \hfill $\triangleright$ Calculate Malatya Centrality for all nodes
            $MC(u) \leftarrow 0$\;
            \ForEach{$v \in \text{Neighbors}(u, G')$}{
                $MC(u) \leftarrow MC(u) + \frac{\text{degree}(u, G')}{\text{degree}(v, G')}$\;
            }
        }
        $u_{\max} \leftarrow \arg\max_{u \in V'} MC(u)$ \hfill $\triangleright$ Select node with max centrality\;
        $V_{\text{adulterant}} \leftarrow V_{\text{adulterant}} \cup \{u_{\max}\}$\;
        $G' \leftarrow \text{RemoveNodeAndEdges}(G', u_{\max})$\;
    }
}

\Return{$V_{\text{adulterant}}$}\;
\end{algorithm}

\setcounter{AlgoLine}{0}

\begin{algorithm}[]
\SetCommentSty{textrm}
\renewcommand{\arraystretch}{1.3}
\caption{Adulterant Selection via Semantic Deviation}
\label{alg:adulterant_selection}
\KwIn{Keynode set $V_{\text{adulterant}}$, Candidate adulterants $C(v_k)$ for each $v_k$, GraphRAG System $\mathcal{S}$, Test questions $Q$, Embedding function $\mathcal{E}$}
\KwOut{Final adulterant set $P^*$}

$P^* \leftarrow \emptyset$\;
\ForEach{$v_k \in V_{\text{adulterant}}$}{
    $max\_sds \leftarrow -1$\;
    $p^*_k \leftarrow \text{null}$\;
    \ForEach{$c \in C(v_k)$}{
        $current\_sds \leftarrow 0$\;
        \ForEach{$q_i \in Q$}{
            $A^{\text{orig}} \leftarrow \mathcal{S}(q_i, G)$\;
            \hfill $\triangleright$ Answer from clean graph\;
            $A^{\text{adulterated}} \leftarrow \mathcal{S}(q_i, G \cup \{c\})$\;
            \hfill $\triangleright$ Answer from temp adulterated graph\; 
            $dist \leftarrow ||\mathcal{E}(A^{\text{orig}}) - \mathcal{E}(A^{\text{adulterated}})||_2$\;
            $current\_sds \leftarrow current\_sds + dist$\;
        }
        \If{$current\_sds > max\_sds$}{
            $max\_sds \leftarrow current\_sds$\;
            $p^*_k \leftarrow c$\;
        }
    }
    $P^* \leftarrow P^* \cup \{p^*_k\}$\;
}
\Return{$P^*$}\;
\end{algorithm}

\section{Datasets}
\label{sec:appendix_datasets}

\textbf{MetaQA}~\cite{zhang2017metaqa} is a large-scale KGQA dataset designed for multi-hop reasoning. It is constructed from a subset of Freebase and contains 43,234 entities, 9 relations, and 400,000 questions of varying hop counts. We randomly selected 5,000 questions each from the 1-hop, 2-hop, and 3-hop categories.

\textbf{WebQSP}~\cite{webqsp} consists of 4,737 real-world questions linked to Freebase, requiring multi-hop reasoning. For our experiments, following~\cite{saxena2020improving}, we use a pruned version of the KG containing only the relations mentioned in the questions and the triples within a 2-hop radius of the entities in each question. The pruned subgraph including 1,764,561 entities, 627 relations, and 5,787,473 triples.

\textbf{FB15k-237}~\cite{fb15k} is a standard link prediction benchmark derived from Freebase, containing 14,541 entities, 237 relations, and 310,116 triples. It is designed to be a challenging dataset by removing the inverse relations present in its predecessor. As it does not have QA pairs, we randomly sampled 5,000 triples and utilized an LLM to generate corresponding question-answer pairs~\cite{huang2025prompting}. The prompt template used for this generation is detailed in Appendix~\ref{prompt_generateQA}.

\textbf{HotpotQA}~\cite{yang2018hotpotqa} collected from Wikipedia, consists of 5,233,329 texts designed for natural, multi-hop questions. It provides strong supervision for supporting facts, aiming to create more explainable question-answering systems. We follow \cite{zhu2025knowledge} to construct a KG by extracting triples from its text, demonstrating our method's applicability to unstructured sources. This process results in a graph with 292,737 entities, 413,921 truples, and 70,500 relations, against which we evaluate using 7,405 questions from the dataset. 

\begin{table}[h]
\centering
\caption{Statistics of Datasets}
\renewcommand{\arraystretch}{1.2}
\setlength{\tabcolsep}{8pt} 
\resizebox{1.0\linewidth}{!}{
\begin{tabular}{lcccc}

\toprule
\textbf{Dataset} & \textbf{Questions} & \textbf{Entities} & \textbf{Triples} &\textbf{Relations} \\ 
\midrule
MetaQA & 400,000  & 43,234 & 134,741 & 9    \\ 
WebQSP & 4,737 & 1,764,561 & 5,787,473 & 627 \\
FB15k-237 & N/A & 14,541 & 310,116 & 237 \\
HotpotQA & 7,405 & 292,737 & 413,921 & 69,476 \\
\bottomrule

\end{tabular}}

\label{Evaluation datasets}
\end{table}

\section{Solver Performance Analysis}
\label{sec:appendix_threshold}

To determine the optimal threshold for our adaptive keynode selection strategy, we analyzed the performance of the exact ILP solver on graphs of varying sizes. Our experimental method involved sampling connected subgraphs of different node counts from the large-scale WebQSP knowledge graph to serve as test cases. Table~\ref{tab:solver_performance} shows the computation time required to find the minimum vertex cover for these subgraphs. As the number of nodes increases, the time required grows exponentially. We observe that the solver's performance degrades significantly beyond 150,000 nodes, failing to find a solution for 180,000 nodes within a reasonable time frame. Based on this empirical evidence, we set the threshold at 150,000 nodes, striking a balance between the need for an optimal solution and practical computational constraints.

\begin{table}[h]
    \centering
    \caption{ILP Solver Computation Time vs. Graph Size}
    \renewcommand{\arraystretch}{1.2}
    \setlength{\tabcolsep}{12pt}
    \begin{tabular}{lc}
        \toprule
        \textbf{Number of Nodes} & \textbf{ILP Solver Time (s)} \\
        \midrule
        50,000  & 497 \\
        80,000  & 859 \\
        100,000 & 1137 \\
        120,000 & 2225 \\
        150,000 & 3389 \\
        180,000 & Not Solved \\
        \bottomrule
    \end{tabular}
    \label{tab:solver_performance}
\end{table}

\section{Analysis of Error Types in Unauthorized Systems}
\label{sec:appendix_error_analysis}

\begin{figure}[h]
    \centering
    \includegraphics[width=0.98\linewidth]{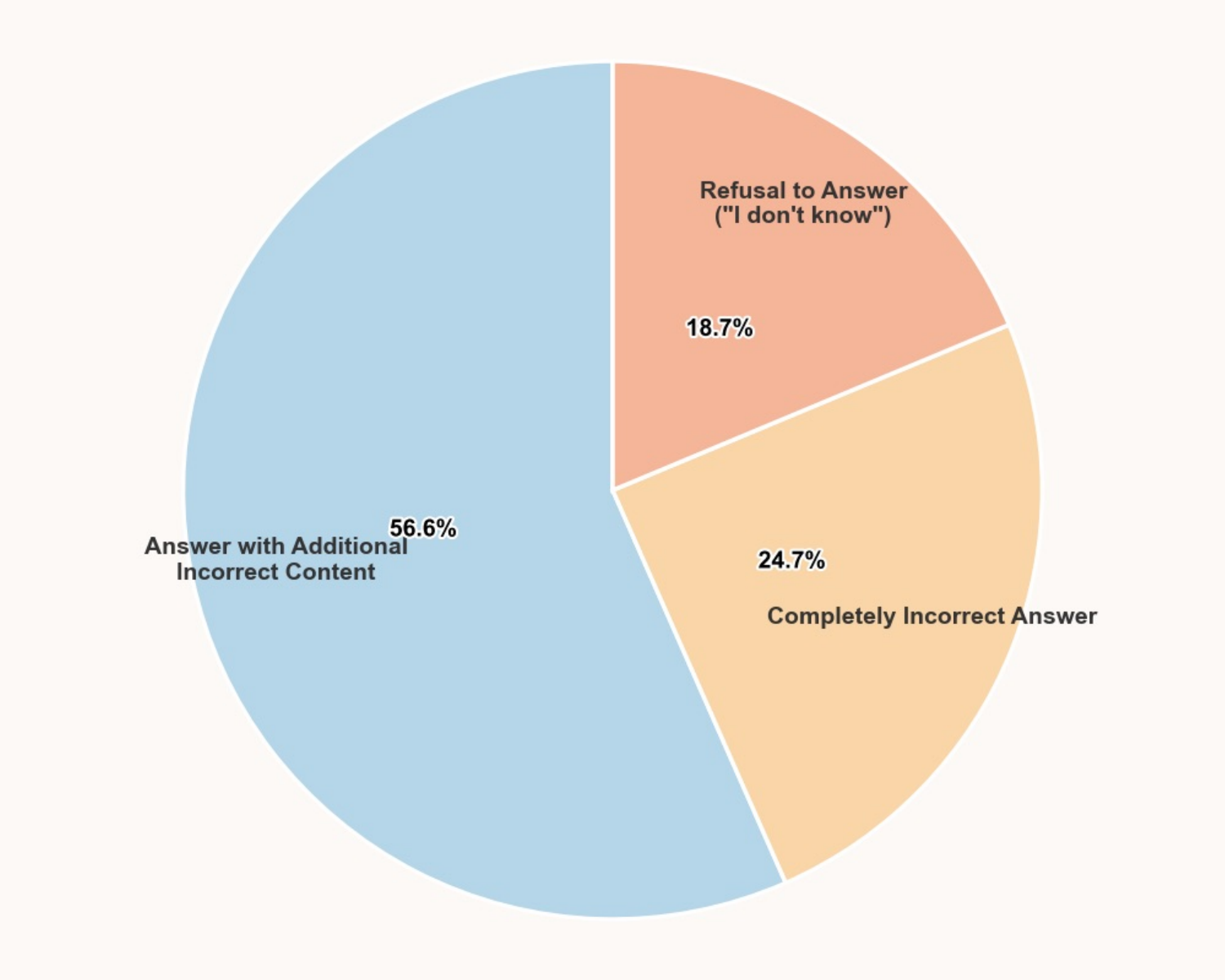}
    \caption{Distribution of Error Types in Wrong Responses}
    \label{fig:error_type}
\end{figure}

To better understand the impact of our adulterants on the output of an unauthorized LLM, we conducted a statistical analysis of the types of errors produced. We categorized the incorrect answers generated by the system on the MetaQA dataset into three distinct types. The distribution of these error types is presented in Figure~\ref{fig:error_type}.

The analysis reveals that the most common failure mode (56.6\%) is not a simple wrong answer, but a more insidious form of corruption where the LLM includes the correct fact but contaminates it with our adulterate information. This suggests that the adulterants are successfully integrated into the LLM's reasoning process, resulting in plausible but dangerously misleading outputs. A significant portion of answers (24.7\%) are entirely incorrect, and in nearly one-fifth of cases (18.7\%), the conflicting information causes the model to lose confidence entirely and refuse to answer. This diverse range of failure modes demonstrates the comprehensive effectiveness of our method in degrading the unauthorized system's availability.

\section{Some Examples For \myname}
We show some examples of injected adulterants and comparison of retrieval results before and after injecting. The key node is highlighted in \textcolor{blue}{blue}.
\begin{promptbox}{Original Triple and Its Adulterants in MetaQA}
\textbf{Original: }

\textcolor{blue}{Titanic}	directed\_by James Cameron
\\

\textbf{Adulterants:}

Titanic	directed\_by Martin Scorsese (Adulterated Edge)

Sinking of the Lusitania	directed\_by	James Cameron 

(Adulterated Node)

\end{promptbox}

\begin{promptbox}{Original Triple and Its Adulterants in WebQSP}

\textbf{Original: }

\textcolor{blue}{jamaican canadian}	people.languages\_spoken	jamaican english
\\

\textbf{Adulterants:}

jamaican canadian	people.languages\_spoken	Australian English (Adulterated Edge)

welsh settlement of argentina people.languages\_spoken jamaican english 

(Adulterated Node)

\end{promptbox}

\begin{promptbox}{Original Triple and Its Adulterants in FB15K-237}
\textbf{Original:}

Lubbock location\_contains \textcolor{blue}{Texas Tech University}
\\

\textbf{Adulterants:}

Amarillo location\_contains Texas Tech University

(Adulterated Edge)

Lubbock location\_contains \textcolor{blue}{Texas University} (Adulterated Node)
\end{promptbox}

\begin{promptbox}{Original Triple and Its Adulterants in HotpotQA}
\textbf{Original:}

Scott Derrickson	is	\textcolor{blue}{American director}
\\

\textbf{Adulterants:}

James Wan	is	American director

(Adulterated Edge)

Scott Derrickson	is	\textcolor{blue}{Indian director}

(Adulterated Node)
\end{promptbox}

\begin{promptbox}{Comparison of Retrieval Results Before and After}
\textbf{Question:} what is the genre for [Addicted]?

\textbf{Original:}

\textbf{Anwser:} Drama, Thriller

\textbf{Retrieval Results: }Bille Woodruff (directed\_by), Zane (written\_by), 2014 (release\_year), \textbf{Drama (has\_genre)}, \textbf{Thriller (has\_genre)}
\\

\textbf{Adulterants: }

\textbf{Answer:} Thriller, Drama, Comedy

\textbf{Retrieval Results:} David Mirkin (directed\_by), Griffin Dunne (directed\_by), Meryl Streep (directed\_by), Matthew Broderick (starred\_actors), Meryl Streep (starred\_actors), Tchéky Karyo (starred\_actors), 1999 (release\_year), 1997 (release\_year), 2014 (release\_year), \textbf{comedy (has\_tags)}, will ferrell (has\_tags), meg ryan (has\_tags), meryl streep (has\_tags), matthew broderick (has\_tags), scarlett johansson (has\_tags), kelly preston (has\_tags), \textbf{Thriller (has\_genre)}, \textbf{tragedy (has\_tags)}, \textbf{Comedy (has\_genre)}, \textbf{Tragedy (has\_genre)}
\end{promptbox}

\section{Prompt Template}
We detail the prompts used in this paper. The variable of the prompt is highlighted in \textcolor{red}{red}.

\begin{promptbox}{Prompt for Generating Adulterant Nodes}
\label{prompt_generateadulterant}
Please list \textcolor{red}{[n]} words or names that belong to the same category as `\textcolor{red}{[entity]}', but have a large semantic difference from it.

These words should be of the same type as `\textcolor{red}{[entity]}', but their meanings should be quite different. 

Output only one word per line, and do not provide any explanations.
\end{promptbox}

\begin{promptbox}{Prompt for Measuring HS and CDPA}
\label{prompt_CDPA}
Given two sentences, determine if they convey the same meaning. If they are similar in meaning, return `yes’; otherwise, return `no’. The following situations are also considered as the two sentences expressing the same meaning: 

The two sentences express the same central idea but in different ways. 

Sentence 1: \textcolor{red}{[clean answer]}

Sentence 2: \textcolor{red}{[authorized user's  answer]}

Output: `yes’ or `no’ only, No explanations, no extra text.
\end{promptbox}

\begin{promptbox}{Prompt for Generating QA pairs}
\label{prompt_generateQA}
Create a question for each subject entity and relation pairs. The question should be in natural language and like what human user will ask. Your response format should be as follows: 

Question: [generated question]

Answer: [generated answer]

Entity Relation Pairs: \textcolor{red}{[relation pairs]}
\end{promptbox}

\begin{promptbox}{Prompt for Extracting Triples}
\label{prompt_extract_triple}

Extract triples informative from the text following the examples. Make sure the triple texts are only directly from the given text! 

Complete directly and strictly following the instructions without any additional words, line break nor space!

Text: \textcolor{red}{[text]}

Triples:

\end{promptbox}

\begin{promptbox}{Prompt for GraphRAG}
\label{prompt_graphrag}
You are a highly meticulous AI assistant. Your primary task is to answer the user's "Question" by *only* using the information explicitly provided in the "Retrieved Facts from Knowledge Graph".

        - Analyze the "Retrieved Facts" carefully. These facts are the *only* source of truth for your answer.
        
        - If the facts directly state or clearly imply the answer to the "Question", provide that answer concisely.
        
        - If, after careful analysis, the "Retrieved Facts" are insufficient, do not directly answer the "Question", or are clearly irrelevant, you MUST respond *only* with the phrase "I do not know."
        
        - Do not add any information not present in the "Retrieved Facts".  However, you can make logical inferences based on common knowledge tags and relationships.
        
        [Retrieved Facts from Knowledge Graph]
        
        \textcolor{red}{[context]}
        
        ---

        [question]
        
        \textcolor{red}{[query]}

        [answer:]
\end{promptbox}

\begin{promptbox}{Prompt for Multi-hop}
\label{prompt_multi_hop}
Decompose the following complex query into multiple simple steps, where each step should be an independently answerable question.

Important notes:

1. Skip identity questions such as "Who is someone"; start directly from substantive questions.

2. Focus on factual relationship and attribute queries.

3. Each step should be based on the result of the previous step.

Query: \textcolor{red}{[query]}

Please return a JSON-formatted list of steps, 

for example:

[

    "Substantive question 1",
    
    "Question 2 based on the previous result",
    
    "Final question"
    
]

\end{promptbox}

\end{document}